\documentclass[twocolumn]{aa}

\usepackage{graphicx} 
\usepackage{times}
\usepackage{amsmath}
\usepackage{textcomp}
\usepackage{amssymb}
\usepackage{lineno}
\usepackage{gensymb}
\usepackage{hyperref}
\usepackage{url}
\usepackage{afterpage}
\usepackage{float}
\usepackage{mathrsfs}
\usepackage{rotating}
\usepackage{csquotes}
\usepackage{graphicx}
\usepackage[caption=false]{subfig}
\usepackage{booktabs}
\usepackage{orcidlink}
\usepackage{threeparttable} 

\newcommand{\nustar}{{\it NuSTAR }}
\newcommand{\xmm}{{\it XMM-Newton }}

\titlerunning{XMM-Newton \& NuSTAR ToO PKS~1725+123}
\authorrunning{Marcotulli et al.}

\begin{document}

\title{{\large XMM-Newton and NuSTAR quasi-simultaneous follow-up of the very high energy flare of PKS~1725+123 in August 2025 }}

\author{L. Marcotulli\orcidlink{0000-0002-8472-3649}\inst{1,2}\corrauth{lea.marcotulli@desy.de}        
     \and A. Gokus\orcidlink{0000-0002-5726-5216}\inst{1}\email{}
     \and L. Heckmann\orcidlink{0000-0002-6653-8407}\inst{3}\email{}
     \and E. Visentin\inst{4}\email{}
     \and I. Jaroschewski\orcidlink{0000-0001-5180-2845}\inst{5}\email{}
     \and J. Otero-Santos\orcidlink{0000-0002-4241-5875}\inst{6}\email{}
     \and M. B\"ottcher\orcidlink{0000-0002-8434-5692}\inst{7}\email{}
     \and M. Zacharias\orcidlink{0000-0001-5801-3945}\inst{8,7}\email{}
     \and R. Takeishi\orcidlink{0000-0001-6335-5317}\inst{9}\email{}
     \and S. Nozaki\inst{9}\email{}
     \and Y. Suda\orcidlink{0000-0002-2692-5891}\inst{10}\email{}
     \and I. Sadeh\orcidlink{0000-0003-1387-8915}\inst{1}\email{}
     \and P. Boorman\orcidlink{0000-0001-9379-4716}\inst{11}\email{} 
     \and E. de O{\~n}a Wilhelmi\orcidlink{0000-0002-5401-0744}\inst{1}\email{} 
     \and D. Berge\orcidlink{0000-0002-2918-1824}\inst{1}\email{}      
     }
\institute{Deutsches Elektronen-Synchrotron DESY, Platanenallee 6, 15738 Zeuthen, Germany
\and Department of Physics and Astronomy, Clemson University, Kinard Lab of Physics, Clemson, SC 29634-0978, USA
\and Universit\'e Paris Cit\'e, CNRS, Astroparticule et Cosmologie, F-75013 Paris, France
\and INFN MAGIC Group: INFN Sezione di Torino and Università degli Studi
di Torino, I-10125 Torino, Italy
\and Universit\'e Paris-Saclay – CEA – Irfu, F-91191 Gif-sur-Yvette, France
\and Istituto Nazionale di Fisica Nucleare, Sezione di Padova, 35131 Padova, Italy
\and Centre for Space Research, North-West University, Potchefstroom, 2520, South Africa
\and Landessternwarte, Universität Heidelberg, Königstuhl, D-69117 Heidelberg, Germany
\and Institute for Cosmic Ray Research, University of Tokyo, 5-1-5, Kashiwa-no-ha, Kashiwa, Chiba 277-8582, Japan
\and Physics Program, Graduate School of Advanced Science and Engineering, Hiroshima University, 1-3-1 Kagamiyama,
Higashihiroshima, 739-8526, Hiroshima, Japan
\and Max Planck Institute for Extraterrestrial Physics, Giessenbachstrasse, 85741 Garching, Germany
}
\date{Received October 30, 2026}

\abstract
{We report the results of the spectral timing analysis of the quasi-simultaneous X-ray follow-up with \xmm ($50\,\rm ks$) and \nustar ($180\,\rm ks$) of the very high energy flare of the blazar PKS 1725+123 ($z=0.586$) in August of 2025. The X-ray monitoring started $\sim3$ days after the announcement of the VHE flare by LST-1, MAGIC and H.E.S.S.~and lasted for almost 5 days. The high sensitivity of \xmm reveals two spectral components, a soft ($\Gamma_{\rm X}>2$) and a hard ($\Gamma_{\rm X}<2$) power law. While the soft component dominates at $E<2\,\rm keV$, the hard component dominates at $E>2\,\rm keV$, likely marking the transition between the synchrotron and the high-energy emission, the ``valley'' of this blazar's spectral energy distribution. Within the \xmm pointing ($\sim14$ hours) we witness the shift of the synchrotron component out of the X-ray band. In the following $\sim 4$ days, \nustar further reveals the brightening by a factor of 1.5 of the hard component, following the shift of the high-energy part of the SED into the X-ray band. Overall, within a week (observed frame) of the VHE flare, PKS 1725+123 went from temporarily being an intermediate synchrotron peaked blazar to transitioning back to its `normal' state of a low synchrotron peaked one. This is the first time that the evolution of a VHE flare from a blazar jet has been followed continuously in X-rays over such an extended period and our results highlight the importance of long-term quasi-simultaneous X-ray follow-up of VHE blazar flares.  }

\maketitle

\nolinenumbers 
   
\section{Introduction}

Very-high-energy (VHE; $E>100\,\rm GeV$) photons are produced in some of the most extreme environments in the Universe. In the extragalactic sky, their production is largely associated with accreting supermassive black holes in the centres of Active Galactic Nuclei (AGN) 
launching powerful relativistic jets (e.g.~blazars, radio galaxies, and $\gamma$-ray narrow-line Seyfert 1 galaxies), as well as starburst galaxies and $\gamma$-ray bursts \citep{CTA_2019}. Blazars, AGN with jets aligned towards Earth \citep{Angel_1980,Urry_1995}, constitute about $\sim40\%$ of the currently classified TeV emitters in the TeV catalog \citep{tevcat}. These sources are some the most powerful particle accelerators in the Universe and, notoriously, their emission is remarkably variable, showing orders of magnitude variability at all wavelengths (from radio to $\gamma$-rays) and timescales, from months to just a few hours \citep[see, e.g.,][for a recent review]{Raiteri_2025}. 

Despite more than two decades of multi-wavelength observations, we do not have a single framework capable of providing a consistent description of blazar emission \citep[see, e.g.,][for a review]{bottcher_2019}. Some of the key open questions revolve around the composition of the jet plasma responsible for the multi-wavelength and multi-messenger emission (leptonic, hadronic, or a combination of both), the jet structure and location of the emission region(s), as well as the mechanisms that accelerate particles (with shock acceleration and magnetic reconnection currently being the most favored scenarios). Characterizing the intensity, shape and timing of a flare from a blazar jet can give us insights into these puzzles. However, the $\gamma$-ray signal from a blazar jet at $E>10\,\rm GeV$ is further shaped by absorption processes, including intrinsic photon-photon interactions within the jet itself and attenuation by the extragalactic background light (EBL, e.g.~\citealp{Dwek_2013}). Therefore, constraints on the particle distribution at these energies suffer from several modeling uncertainties. Additionally, follow-up multi-wavelength observations of these flares are often non-simultaneous, making the physical interpretation of flares and their evolution especially challenging.

Importantly, simultaneous X-ray and VHE observations of blazar flares are crucial. Indeed, X-ray coverage of a VHE blazar flare can constrain the shape and normalization of the blazar’s high-energy spectral energy distribution (SED), providing a powerful test of physical jet models and particle-acceleration scenarios \citep[e.g.][]{hess+21_1510,nievas_rosillo_2025, op_313_2026}. In particular, long duration high-sensitivity broadband X-ray monitoring of these flares can be an exceptional probe of the SED evolution during the flare, which in turn constrains the underlying physical parameter evolution \citep[see][for a review on X-ray blazar science]{Marcotulli_2024}. Simultaneous observations with the high sensitivity of {\it Nuclear Spectroscopic Telescope Array} ({\it NuSTAR}; $3-79\,\rm keV$; \citealp{nustar_2013}), complemented by the soft X-ray anchor with \xmm ($0.2-10\,\rm keV$; \citealp{xmm_2001}) are to date the best approach while awaiting a broadband X-ray spectrometer such as the proposed High-Energy X-ray Probe (HEX-P, \citealp{hexp_overview_2024}) or High EneRgy Observatory for Imaging X-rays (HEROIX\footnote{\url{https://sti.usra.edu/missions/heroix/}}). Broadband $0.2-79\,\rm keV$ coverage of these VHE flares can probe the particle population responsible for the emission, and together with simultaneous TeV observations, these data can offer one of the most stringent tests of blazar emission models to date.

During \nustar AO11, the team secured a Target-of-Opportunity (ToO) program for joint \nustar (180 ks) and \xmm (50 ks) observations to capture a VHE flare from an extragalactic source with simultaneous broadband X-ray coverage (program ID: 11266). A dense multi-wavelength follow-up campaign, spanning the radio-to-optical bands, was also organized to be triggered simultaneously to the VHE flare. We subsequently monitored the $\gamma$-ray sky continuously, awaiting the onset of a suitable flare from an interesting source. 
On the 19th of August 2025, LST-1 (Large-Sized Telescope-1, $20\,\rm GeV - 3\, TeV$) and MAGIC (Major Atmospheric Gamma-ray Imaging Cherenkov, $50\,\rm GeV - 30\, TeV$) followed by H.E.S.S. (High Energy Stereoscopic System, $60\,\rm GeV - 100\, TeV$)  announced the discovery of a flare from a flat spectrum radio quasar (FSRQ) type blazar \citep{atel_magic_lst, atel_hess}: PKS 1725$+$123, with redshift $z=0.586$ \citep{Gorshkov_2003, Healey_2008}, located at $\rm R.A.(\mbox{J2000})=17h~28m~07.0512s$ and $\rm Dec(\mbox{J2000})=+12d~15m~39.485s$ \citep{Xu_2019}. The source was detected with a statistical significance of more than $5\sigma$ after about 2 hours of observations with MAGIC and LST-1, and above $6\sigma$ with H.E.S.S. Based on the uniqueness of the source (see Section~\ref{sec:why}) and the intensity of the VHE flare (private communication with the LST-1, MAGIC and H.E.S.S. Collaboration teams),
we triggered the \xmm and \nustar ToO observations.

Here we present the results on the \xmm and \nustar quasi-simultaneous follow-up of the VHE flare from PKS 1725+123. In Section~\ref{sec:target}, we describe the target selection and X-ray campaign; in Section~\ref{sec:x-ray_ana} and  in Section~\ref{sec:x-ray_fit}, we detail the X-ray spectral-timing analysis and the X-ray fitting, respectively. In Section~\ref{sec:res_disc} we discuss the interpretation of the results and, finally, Section~\ref{sec:outl} reports the conclusions and outlook of this work. 

\section{Target selection and X-ray monitoring campaign}\label{sec:target}

\subsection{Why trigger on PKS 1725+123?}\label{sec:why}
PKS 1725+123 had never been detected before at VHE $\gamma$-rays, though it had already sparked interest in the community due to its potential association with the IceCube event IC-201021A \citep{2020GCN.28715....1I,Nanci_2022}.
On March 20th, 2025, the {\it Fermi}-Large Area Telescope \citep[LAT,][]{LAT_2009} reported an increase in the $E>100\,\rm MeV$ flux of the source by a factor of 30 with respect to the average flux listed in the release of the Fourth {\it Fermi}-LAT catalog (4FGL-DR4, \citealp{4fgl_dr4}); this also corresponded to a hardening of the photon index (from $\Gamma_{\rm 4FGL}=2.47\pm0.05$ to $\Gamma_{\rm flare}=1.9\pm0.2$, \citealp{LAT_march_2025}). The $\gamma$-ray flare was closely followed by the report of a historical maximum of the source in the near-infrared \citep{NIR_2025ATel}, though no significant detection of the source by VHE instruments was recorded. On July 31st, 2025, the source showed renewed $\gamma$-ray activity at a flux level $>30$ times the average 4FGL-DR4 flux and again a hardening of the photon index \citep{LAT_2025ATel}.  Then, on August 19th 2025, the source was announced as discovered at VHE for the first time by LST-1, MAGIC and H.E.S.S. \citep{atel_magic_lst,atel_hess}. This also corresponded to a historical maximum of the optical emission from the source \citep{atel_optical_aug_2025}, and a $\sim 4$ times increase in radio flux in the 7mm band \citep{eff_atel_2025}. 

Besides the fact that this source had just been discovered at VHE, PKS~1725+123 belongs to a blazar class that can only be detected by VHE telescopes if an extreme flare occurs. Typical blazars' SEDs have a double peak shape, with the low-energy component being produced by synchrotron emission of electrons in the jet, and the high-energy component explained via either inverse Compton scattering of the leptons in the jet, or hadronic/lepto-hadronic processes. The majority of blazars detected at VHE are the ones whose synchrotron peak is $\nu_{\rm synch, peak}>10^{14}\,\rm Hz$ and, correspondingly, whose high energy peak falls in the GeV-TeV energy band, so called intermediate (ISP) and high-synchrotron peaked (HSP) sources. They also commonly belong to the BL Lac blazar class, i.e.~their optical spectrum shows no or very narrow emission lines. A very small fraction of VHE emitting blazars belongs to the low-synchrotron peaked (LSP, $\nu_{\rm synch, peak}\leq10^{14}\,\rm Hz$ and high-energy peak in the MeV band) and FSRQ (showing broad emission lines in the optical) categories \citep[e.g.,][]{op_313_2026}. Interestingly, PKS 1725+123 is classified as an LSP FSRQ source, with a synchrotron peak at $\nu_{\rm sync,peak}\sim5.3\times10^{12}\,\rm Hz$ and a high energy peak at $\nu_{\rm HE,peak}\sim 2\times 10^{21} - 5\times10^{22}$ ($E_{\rm HE, peak}=125\pm117\,\rm MeV$ as listed in the Fourth Catalog of AGN Detected by the {\it Fermi}-LAT \citealp[4LAC-DR3,][]{4lac-dr3}). Therefore, for the source to be significantly detected at $E>20\,\rm GeV$ during the VHE flare, meant that something extreme happened in/around the jet to either momentarily energize the radiating particles, significantly increase the photon field density, and/or change the jet physical parameters. 
Moreover, it is the 10th most-distant source ever seen in VHE $\gamma$-rays, only the 12th FSRQ ever detected at VHE, and only the 6th detected at $z>0.55$ \citep{tevcat}.

\subsection{When did the trigger start?}

 Due to difficulties uplinking the pointing commands to \nustar (private communication with the \nustar SOC), together with several other GO ToOs already scheduled for both observatories that week, the \xmm observations were conducted from 2025 August 22 at 05:10:19 UTC to 23:13:41 UTC. The \nustar observations began on 2025 August 22 at 19:31:06 UTC and continued until 2025 August 26 at 16:15:28 UTC. This means that: (1) we could not observe the source strictly simultaneously with the start of the flare at very high-energies, and (2) the overlap between the soft and hard X-ray monitoring is less than four hours. Moreover, the \xmm observations were dominated by solar flares during the last 2 hours of observations (see Section~\ref{sec:xmm_lc}), hence a strictly simultaneous constraint on the $0.2-80\,\rm keV$ spectrum is not possible. Nevertheless, we were able to monitor the evolution of the VHE flare in the X-ray band beginning just three days after its detection, {\it continuously for five consecutive days} using highly sensitive instruments covering the X-ray band. To our knowledge, this is the first time that the decay of a VHE flare from a blazar jet has been observed continuously in X-rays over such an extended period.




\begin{figure}
    \centering
    \includegraphics[width=0.98\linewidth]{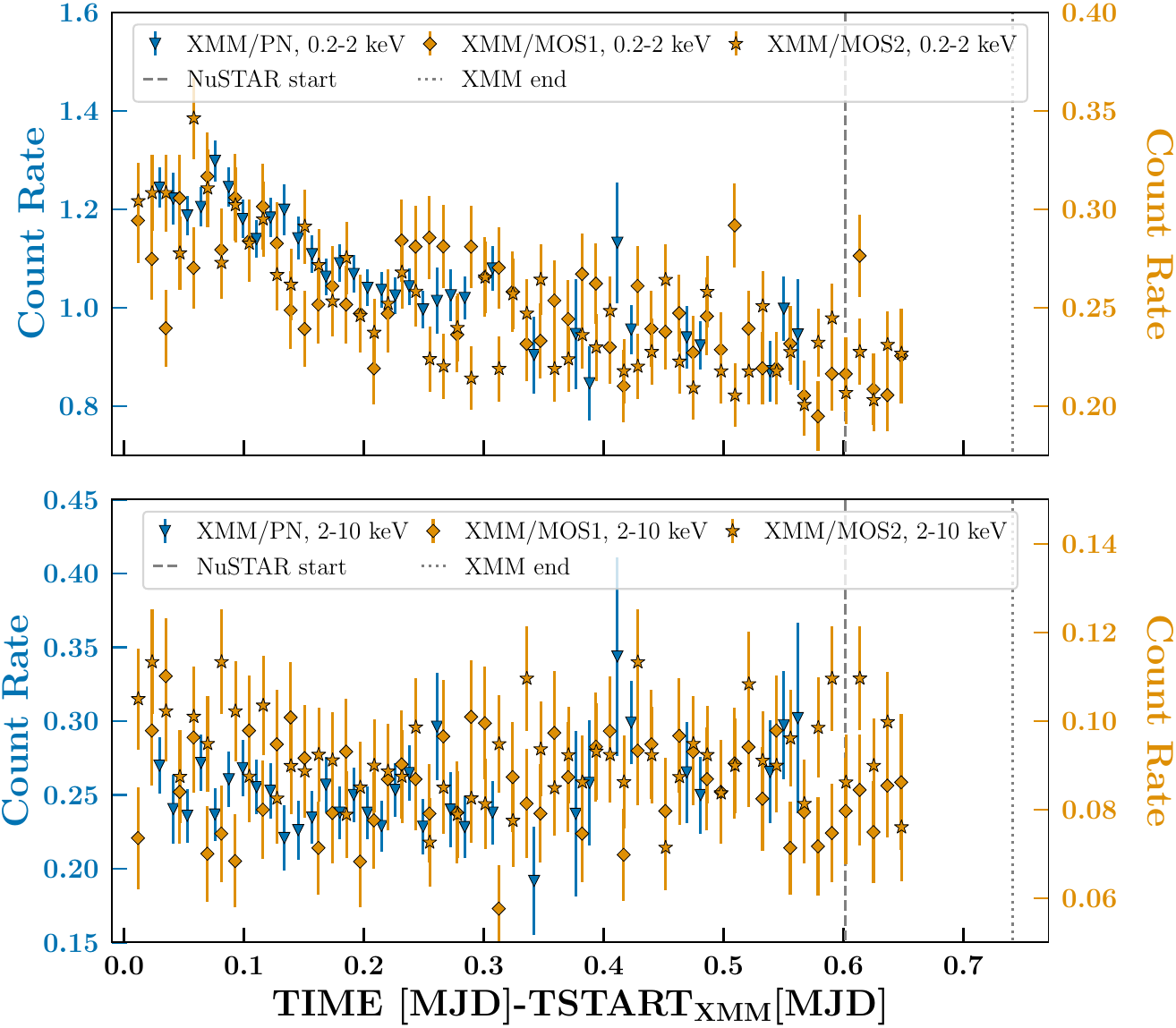}
    \caption{Background subtracted \xmm source count rates obtained using $1\,\rm ks$ time bin as function of time (in MJD from the start of the \xmm observation) for PN (blue triangles), MOS1 (yellow diamonds) and MOS2 (yellow stars) in the $0.2-2\,\rm keV$ (top panel) and the $2-10\,\rm keV$ (bottom panel) band. Because the MOS1 and MOS2 count rates are lower than the PN ones, the plots use a dual y-axis for ease of comparison: the left (blue) y-axis is scaled according to PN count rates, and the right (yellow) one according to MOS1/MOS2. The dashed vertical lines indicate the start of the \nustar observation, while the dotted vertical line marks the end of the \xmm observation.} 
    \label{fig:cr_lc_xmm}
\end{figure}

\begin{figure}
    \centering
    \includegraphics[width=0.98\linewidth]{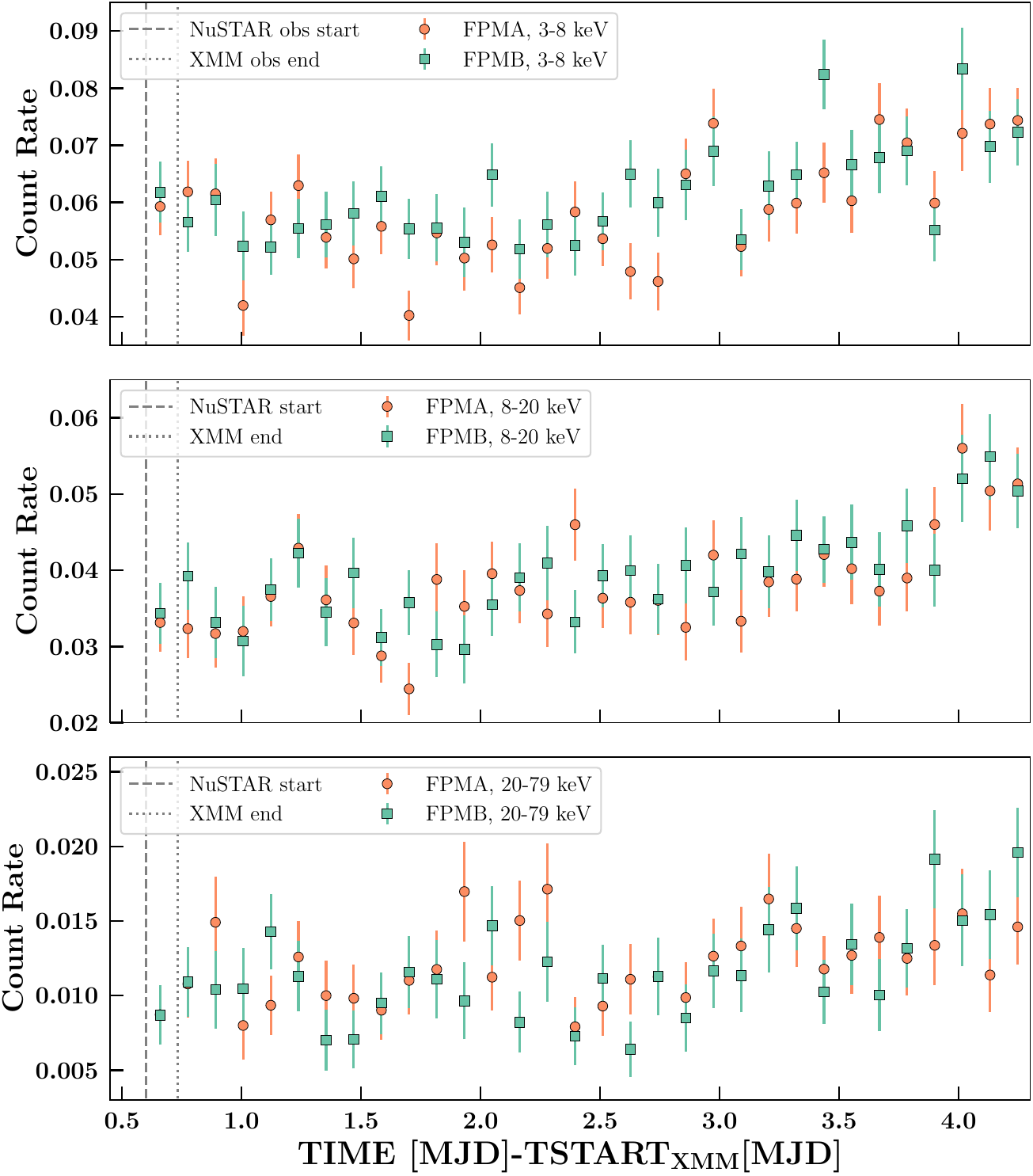}
    \caption{Background subtracted \nustar source count rates as function of time (in MJD from the start of the \xmm observation) binned in $10\,\rm ks$ time bins in the $3-8\,\rm keV$ (top panel), $8-20\,\rm keV$ (middle panel) and $20-79\,\rm keV$ (bottom panel) bands for both FPMA (orange circles) and FPMB (blue/green squares). The dashed vertical line indicates the start of the \nustar observation, while the dotted vertical line marks the end of the \xmm observation.}
    \label{fig:cr_lc_nu}
\end{figure}

\section{X-ray Analysis}\label{sec:x-ray_ana}

\subsection{XMM-Newton}
\subsubsection{Spectrum generation}\label{sec:xmm_spec}
The \xmm spectra were extracted following the standard reduction pipeline using the Science Analysis System (SAS\footnote{\url{https://www.cosmos.esa.int/web/xmm-newton/sas}}) v22.1.0 and the HEASoft v6.35 software packages:
\begin{enumerate}
    \item MOS1 and MOS2 raw files were processed with the \texttt{emproc} task, and the PN ones with \texttt{epproc}; 
    \item we removed bad pixels from MOS1/MOS2 and PN observations using \texttt{evtselect} selecting  ``(PATTERN $\leq$ 12) \&\&(FLAG==0)"" for MOS1/MOS2 and ``(PATTERN $\leq$ 4) \&\&(FLAG==0)" for PN;
    \item after generating the high energy ($>10\,\rm keV$) light-curves of the observation, we decided to select times with count rates below $0.3\,\rm cts/s$ in the high energy light-curves for MOS1 and MOS2 observations, and below $0.2\,\rm cts/s$ from PN. These good time intervals (GTIs) were then applied to the event files;
    \item to generate the source and background regions, we used SAOImageDS9 \citep{Joye_2003, Fruscione_2026}. The source region was defined as a circle of $30''$ located at the source optical position and then centroided on the X-ray counts. The background region was also defined as a circle of $30''$ located close to the source but away from source contribution;
    \item then, we generated the spectra for both the source and the background regions using \texttt{evtselect};
    \item the \texttt{backscale} function was used to calculate the area of the source and background regions, and to account for chip gaps and bad pixels in the spectra; 
    \item the ARF and RMF files were generated with the \texttt{rmfgen} and \texttt{arfgen} tasks, and then were added to the source and background spectrum;
    \item finally, we grouped the source spectra using the optimal binning scheme of \citet{Kaastra_2016} available via \texttt{ftgrouppha}.
\end{enumerate}

\subsubsection{Light-curve generation}\label{sec:xmm_lc}
To generate the \xmm {\it count rate} light-curves in search for variability, we also followed common procedures:
\begin{enumerate}
    \item we first applied the baryocentric correction to the cleaned event file (created in step 3 of Section~\ref{sec:xmm_spec}) using \texttt{barycen};
    \item we then produced the light-curves of the source and background regions (defined in step 4 of Section~\ref{sec:xmm_spec}) in the desired energy ranges with a time bin of $1\,\rm ks$;
    \item finally we subtract background and artifacts from the source light-curve using \texttt{epiclccorr}.\
\end{enumerate}

We generated {\it count rate} light-curves both in the soft ($0.2-2\,\rm keV$) and hard ($2-10\,\rm keV$) X-ray band to check for obvious variability behaviors of the source during the long term monitoring. The {\it count rate} light-curves are shown in Figure~\ref{fig:cr_lc_xmm}. As can be seen, the soft X-ray count rates decrease with time for all three instruments (MOS1, MOS2 and PN), while the hard X-ray ones remain constant. We note that the \xmm observations were severely affected by solar flares, specifically towards the end of the observation. In particular, PN data are unusable past MJD 60909.796 ($\sim70\%$ of the observation is usable) and MOS1/MOS2 past MJD 60909.883 ($\sim87\%$ of the observation is usable).

\subsection{NuSTAR}\label{sec:nu_spec}

The \nustar spectra and {\it count rate} light-curves were extracted by following the standard reduction pipeline using the HEASoft v6.35 software package and the Calibration Database (CALDB) files v20260630: 
\begin{enumerate}
    \item the raw files were processed with the \texttt{nupipeline} task;
    \item we defined a source region as a circle of $35''$ located at the source optical position and then centroided on the X-ray counts using DS9. The background region was also defined as a circle but with a size of $60''$, located close to the source but away from source contribution;
    \item as advised by the \nustar SOC, we first defined GTIs to remove the times in our observations affected by solar flares\footnote{\url{https://github.com/NuSTAR/nustar-gen-utils/blob/main/notebooks/GTI_filter_solar_flare.ipynb}};
    \item we then extracted the source and background spectra and their respective light-curves using the \texttt{nuproducts} task with our user defined GTI, for both focal plane modules (FMP) A and B;
    \item finally, we grouped the source spectra using the optimal binning scheme of \citet{Kaastra_2016} available via \texttt{ftgrouppha}. 
\end{enumerate}

To check for variability in the {\it count rates}, we produced light-curves in the $3-8\,\rm keV$, $8-20\,\rm keV$ and $20-79\,\rm keV$ ranges using a $10\,\rm ks$ time bins. The light-curves are shown in Figure~\ref{fig:cr_lc_nu} (right panel). Visually, we can see how the count rates increase  towards the end of the observations. 

\begin{figure*}
    \centering
    \includegraphics[width=0.33\linewidth]{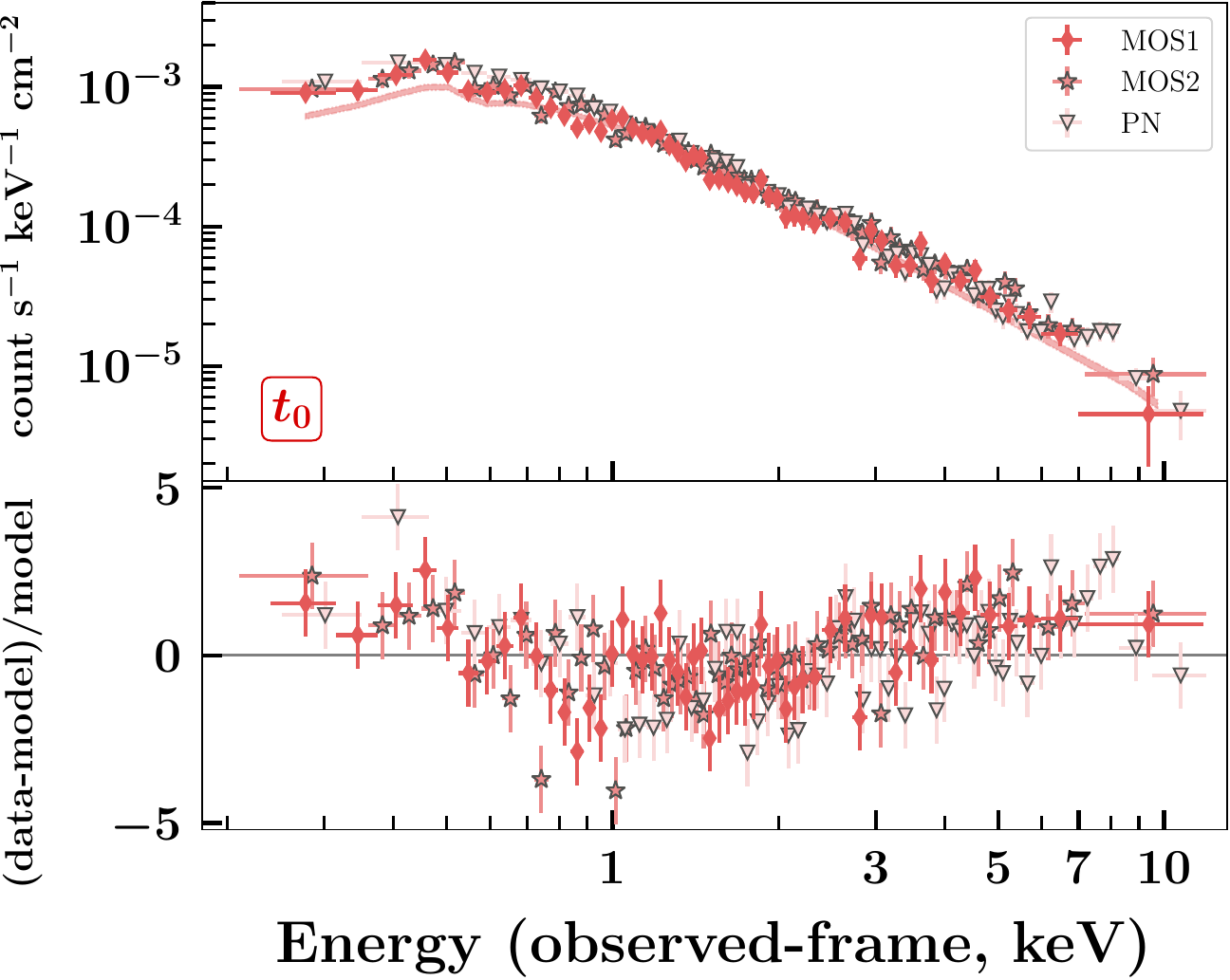}
    \includegraphics[width=0.33\linewidth]{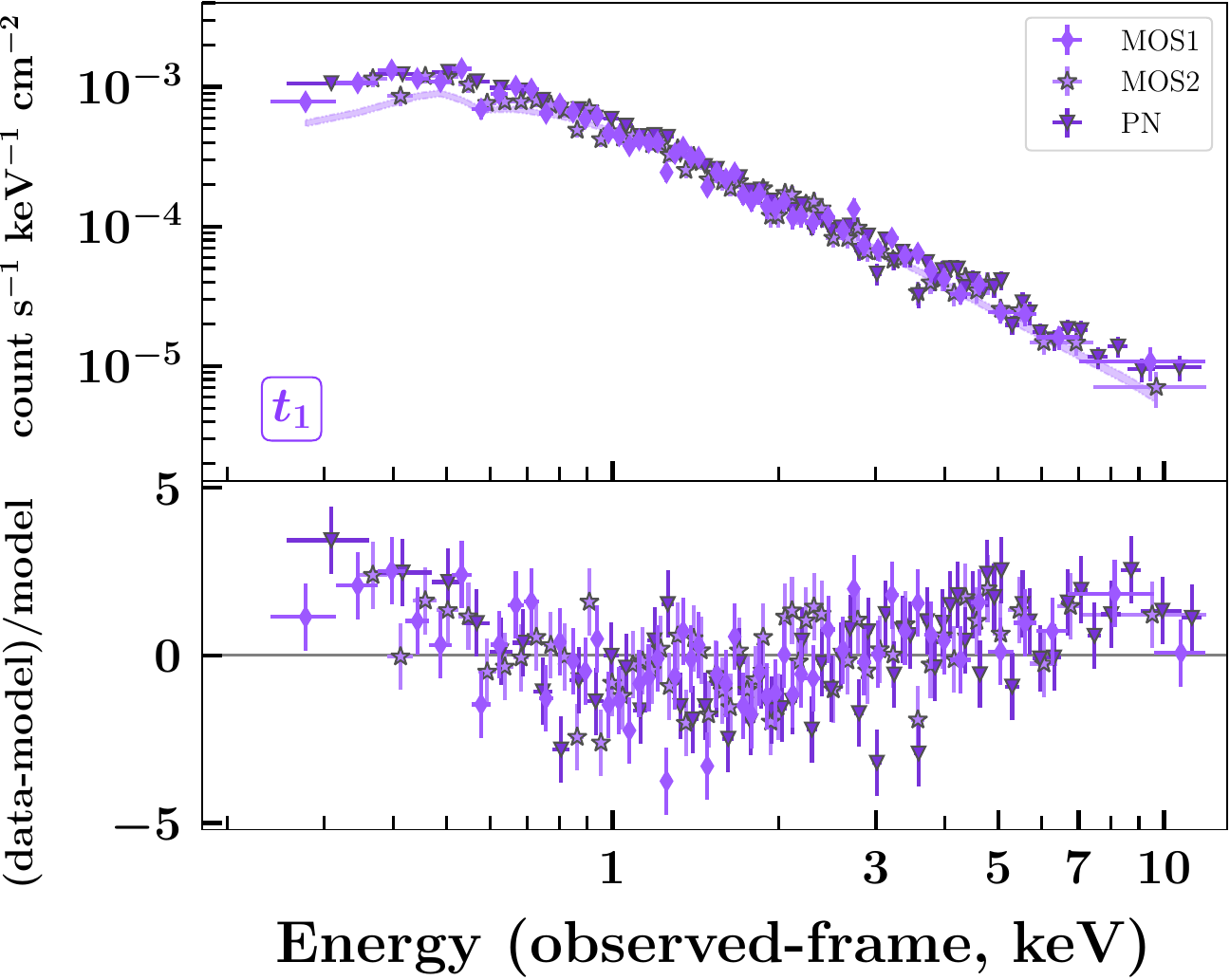}
    \includegraphics[width=0.33\linewidth]{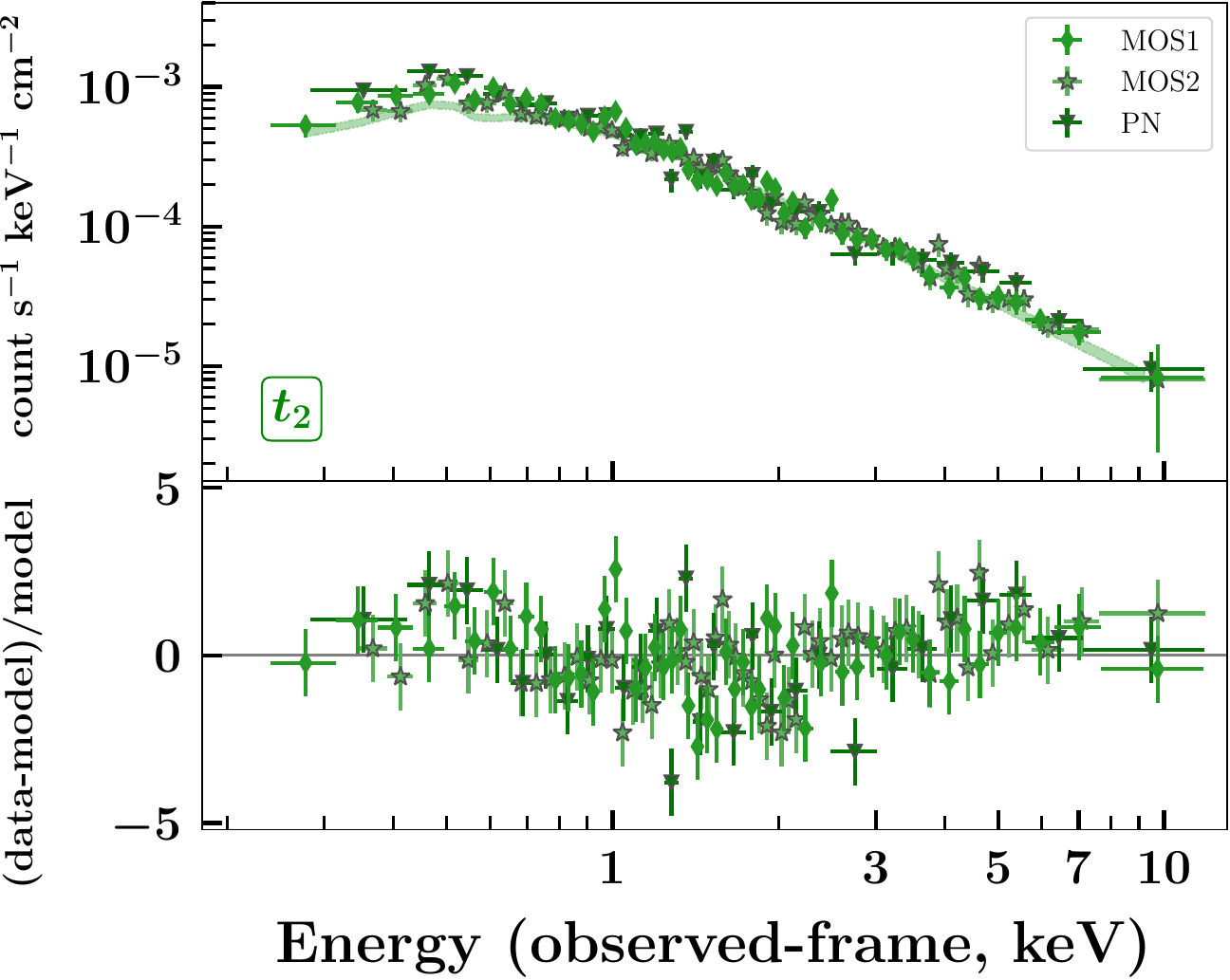}
    \includegraphics[width=0.33\linewidth]{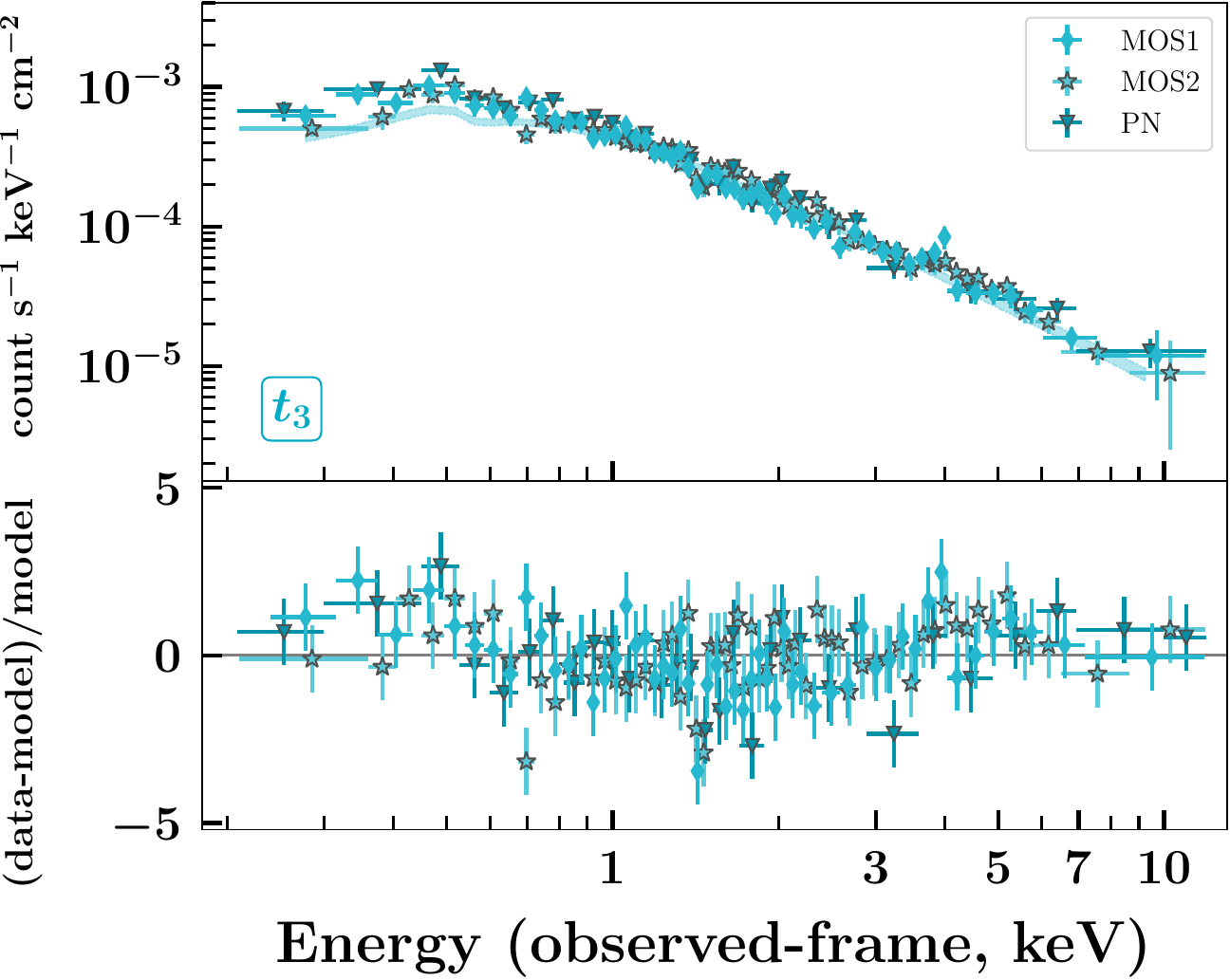}
    \includegraphics[width=0.33\linewidth]{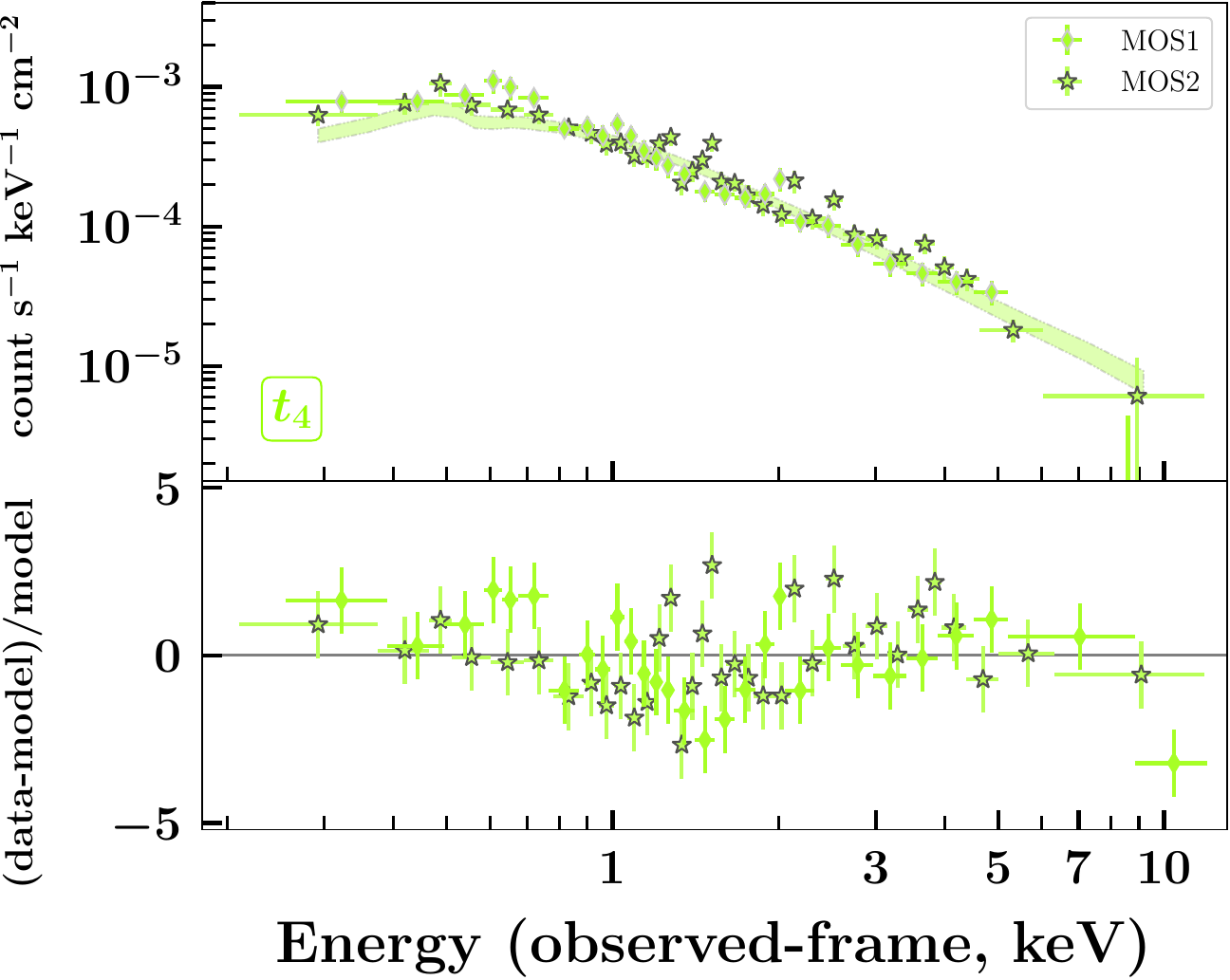}
    \caption{Background-subtracted \xmm count rate spectra and residuals, after binning each spectrum to have a minimum significance of five per bin for visual clarity. The source counts are normalized by the average effective area in each bin. The plots, from left to right, and top to bottom, are ordered as function of time bin, from $t_0$ to $t_4$ (see Table~\ref{tab:time_bins}) and the data points are color coded from red ($t_0$) to light green ($t_5$). The MOS1, MOS2 and PN data are marked with diamonds, stars and triangles, respectively. In each panel, we show the model predictions for the redshifted single power-law fits to the data, with the shaded areas (colored-coded like the data points) encompassing the 68th percentile range region of the fluxes posterior distribution.}
    \label{fig:xmm_fit_pl}
\end{figure*}

\subsection{GTIs for spectral timing analysis}

The observation of the source lasted for a total of $\sim 5$ days between \xmm and \nustar. As the count rate light-curves of both instruments show a hint of variability both in the soft X-ray and the hard X-ray band, we wanted to sample possible spectral evolution of the source as the flare progressed. The elevated X-ray state of the source after the VHE flare ensures that the signal-to-noise ratio of our observations is high enough to produce time-resolved spectra. 

To do so, we selected $\Delta t_{\rm XMM}=10\,\rm ks$ time bins for \xmm and a $\Delta t_{\rm NuSTAR}=40\,\rm ks$ time bins for \nustar (which effectively amounts for about 20 ks of integrated exposure time per bin accounting for gaps in the observation). The time bins for both instruments are listed in Table~\ref{tab:time_bins} of Appendix~\ref{app:a}.  After creating the GTIs, we repeated the spectral generation steps listed above (Sections~\ref{sec:xmm_spec} and \ref{sec:nu_spec}) including the time intervals and we extracted the time-resolved source spectra.

\begin{figure*}
    \centering
    \includegraphics[width=0.33\linewidth]{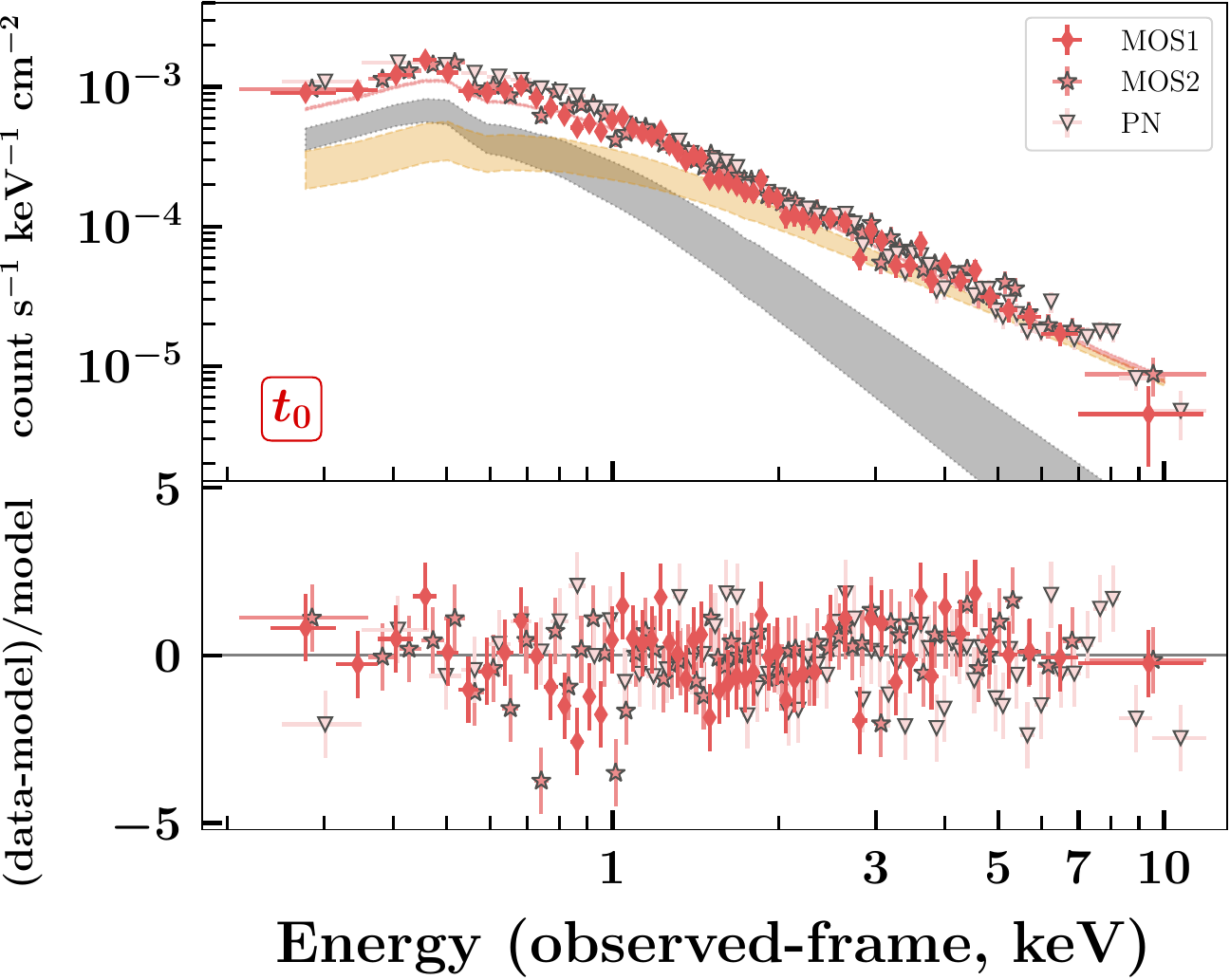}
    \includegraphics[width=0.33\linewidth]{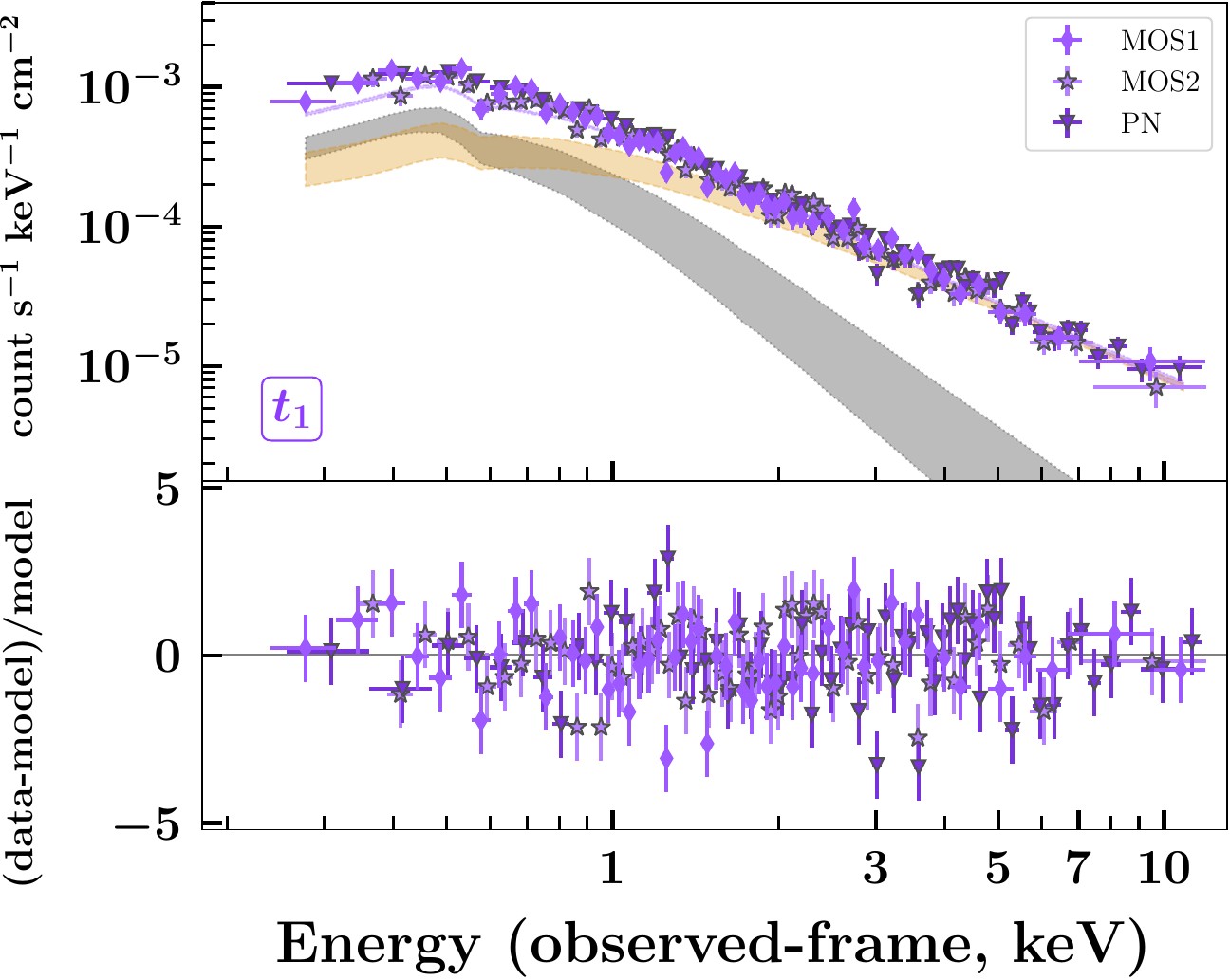}
    \includegraphics[width=0.33\linewidth]{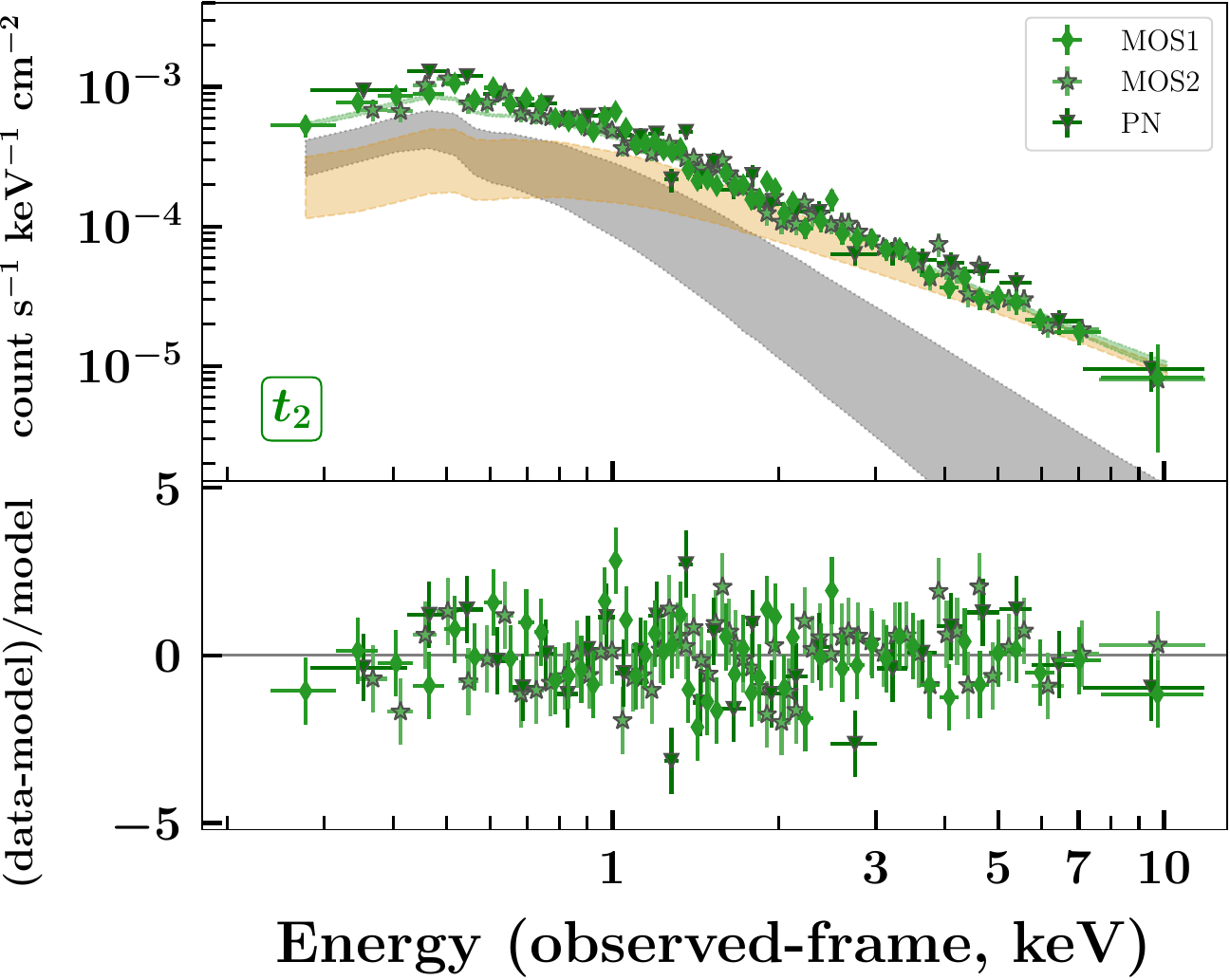}
    \includegraphics[width=0.33\linewidth]{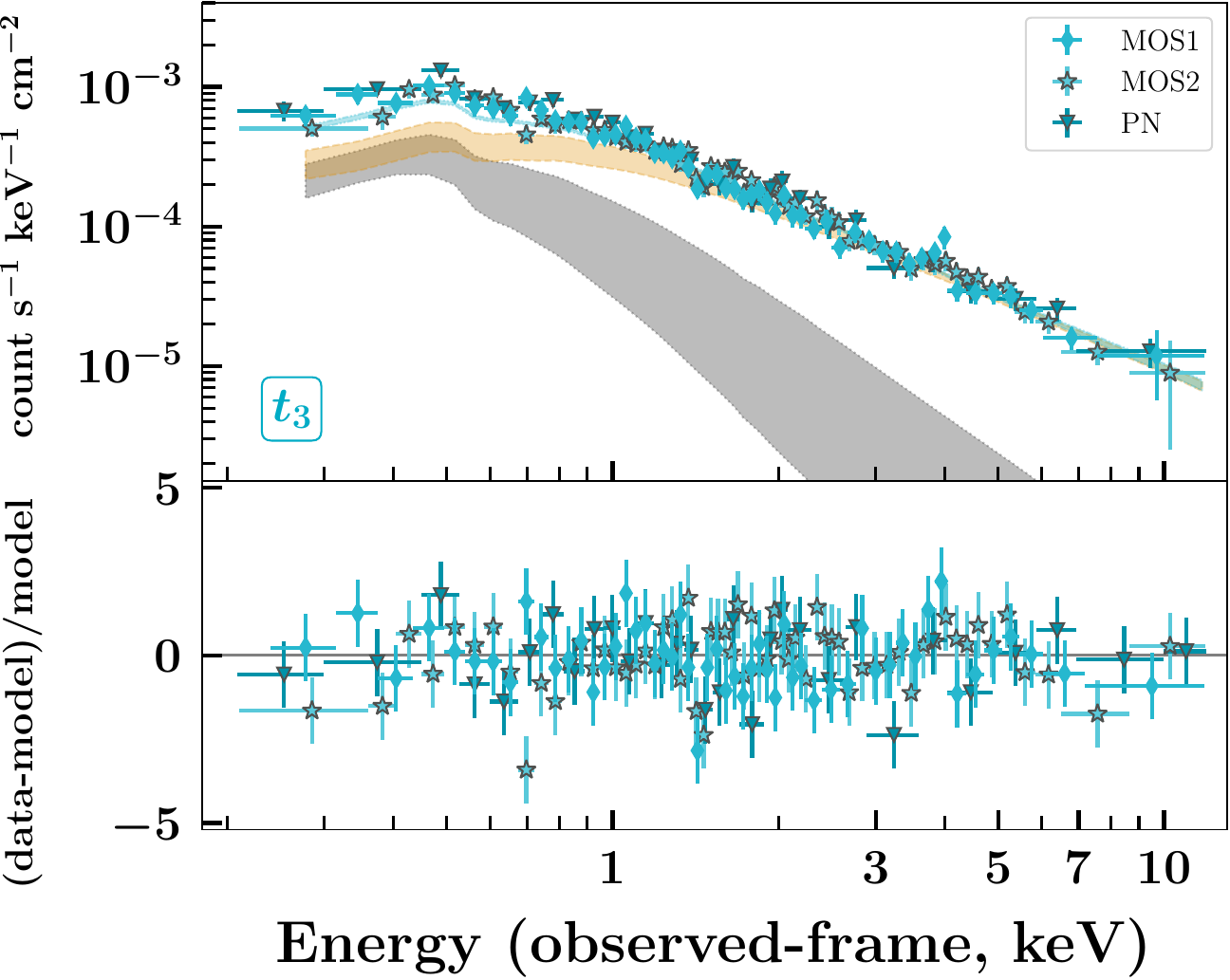}
    \includegraphics[width=0.33\linewidth]{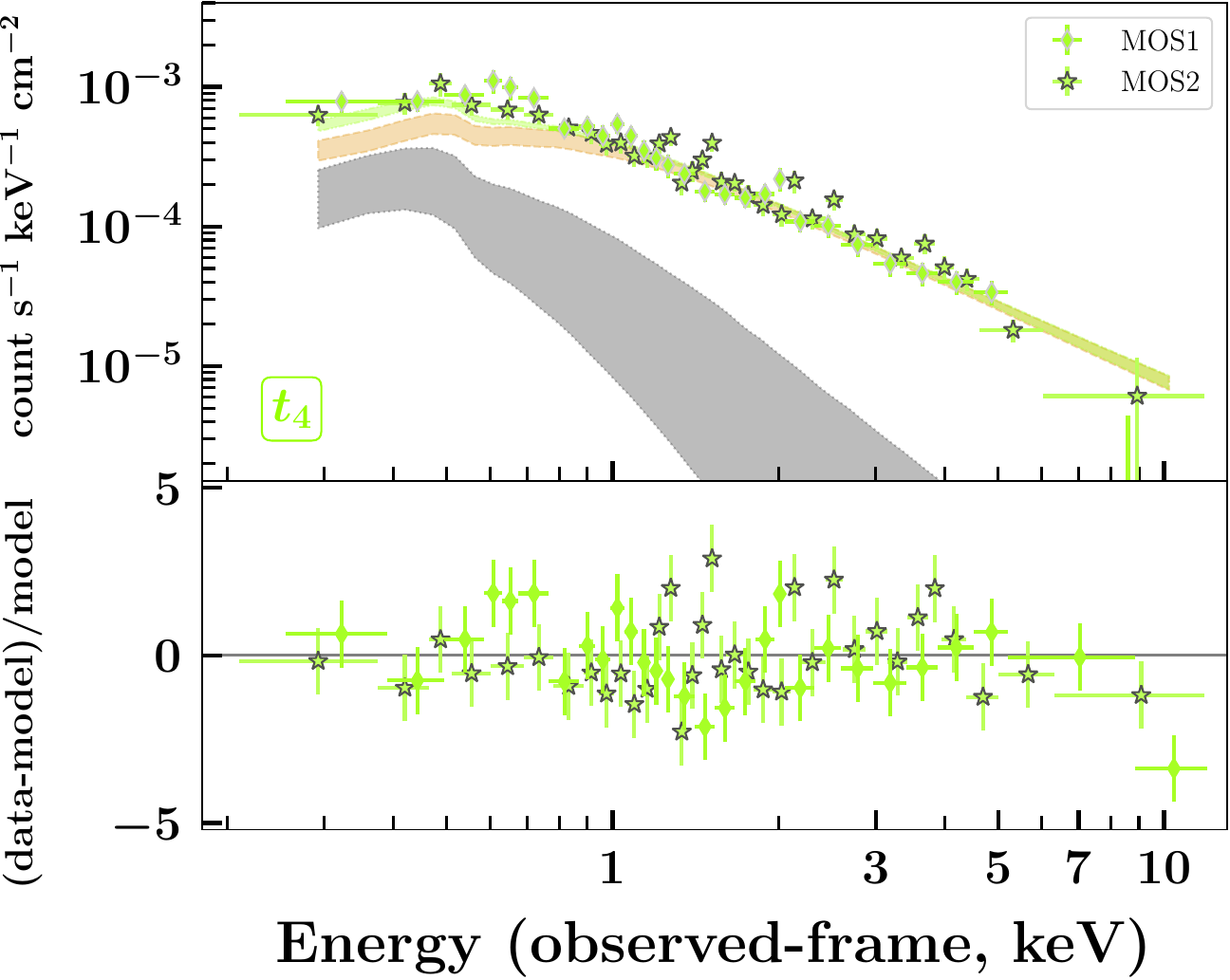}
    \caption{Background-subtracted \xmm count rate spectra and residuals, presented using the same binning, normalization, time-bin ordering, colors, and symbols as in Figure~\ref{fig:xmm_fit_pl}. 
    In each panel, we show the model predictions for the redshifted double power-law fits to the data, with the shaded areas encompassing the 68th percentile range region of the fluxes posterior distribution. The gray shaded area represent the ``soft component" contribution, the pink shaded area the ``hard component", and the one colored like the data is the total emission. }
    \label{fig:xmm_double_pl}
\end{figure*}

\begin{figure*}
    \centering
    \includegraphics[width=0.49\linewidth]{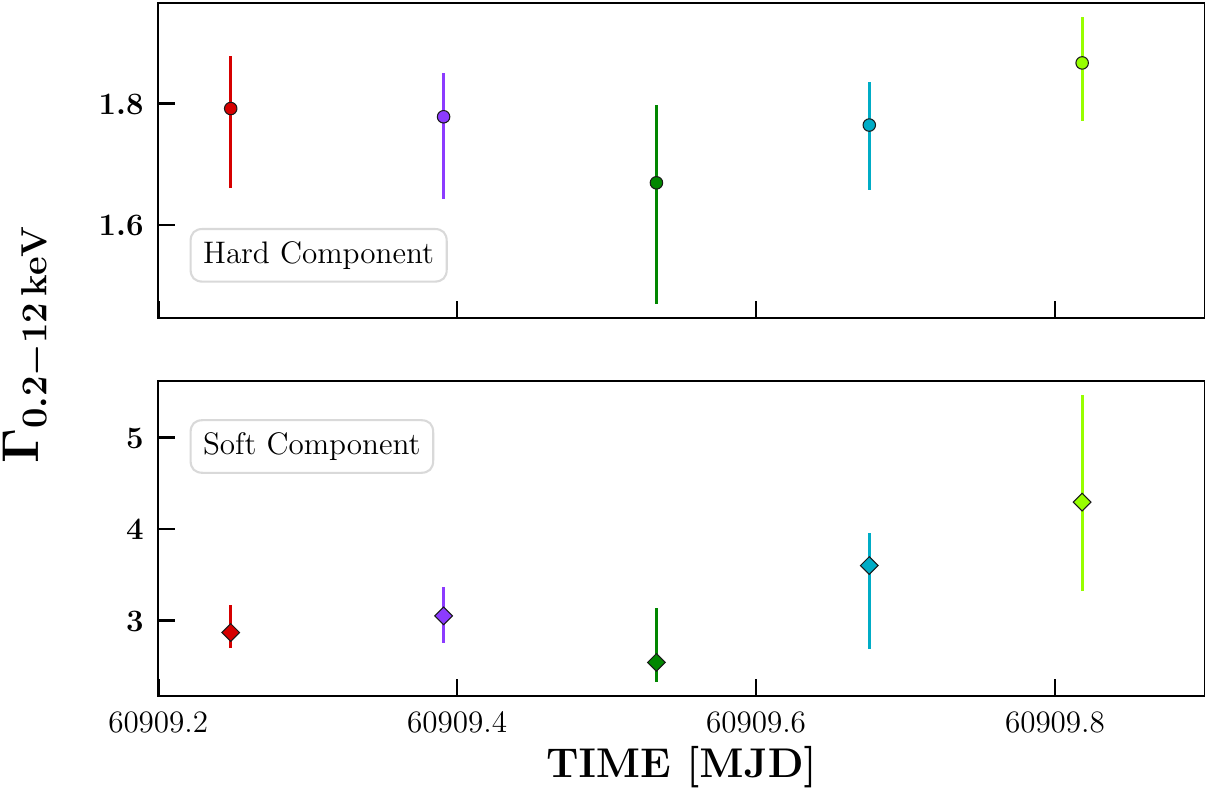}
    \includegraphics[width=0.49\linewidth]{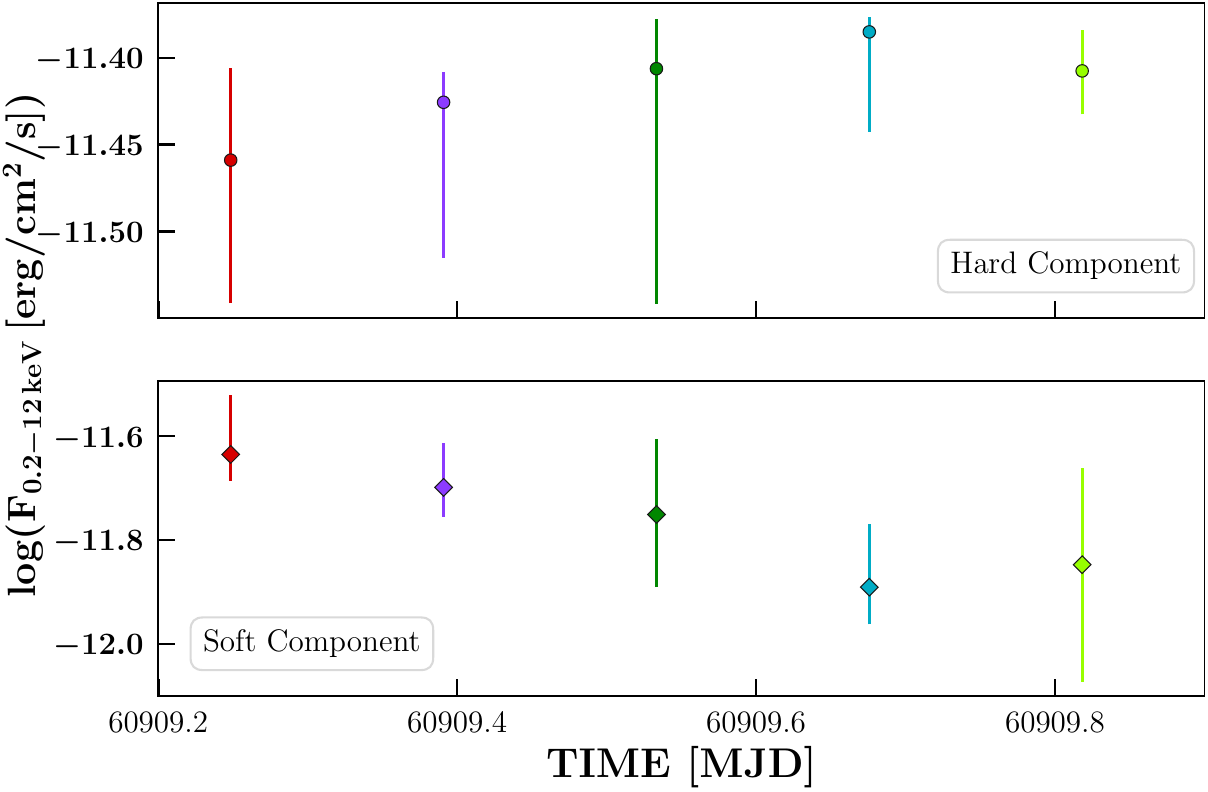}    
    \caption{Temporal evolution of the $0.2 - 12\,\rm keV$ photon indices (left) and fluxes (right) of the two power-law components detected in the \xmm data.}
    \label{fig:xmm_lc_ph_fl}
\end{figure*}

\section{X-ray fitting}\label{sec:x-ray_fit}

\subsection{Fitting with the Bayesian X-ray software}


The X-ray spectral fitting was performed using the Bayesian X-ray Analysis software  (BXA, v.4.1.1, \citealp{Buchner_2014,Buchner_2016}). This tool links the commonly used X-ray SPECtral fitting software (\texttt{XSPEC}, \citealp{1996ASPC..101...17A,Gordon_2021}) to the nested sampling algorithm \texttt{UltraNest} \citep[v. 3.5.7,][]{Buchner_2021}, allowing us to perform a Bayesian fit of the data. We begin by assigning a prior distribution of possible values to each of the fitted parameters. Then, the algorithm iteratively explores the likelihood space across all parameter combinations, discarding those with lower likelihoods. Convergence is reached when the contribution to the Bayesian evidence integral stabilizes and the likelihood, weighted by the discarded volume, becomes negligible. The software returns the full posterior distributions of every fitted parameter. 

Throughout the paper, we quote the mode of the distributions as our ``best-fit'' parameter value, and the 68\% highest density interval (HDI) of the distribution as our uncertainty. We further compute the flux posterior distribution by calculating the flux (in a specific energy band) for every combination of posterior parameters, and then extract the mode and HDI of this distribution. 
The choice of priors depends on the fitted models and free parameters, which we detail in the Sections below for every tested model. In Table~\ref{tab:params_prior} of Appendix~\ref{app:a}, we summarize the allowed range of values for each of the fitted parameters and assigned priors. Throughout, we minimize the modified C-statistic \citep[W-stat;][]{1979ApJ...228..939C,1979ApJ...230..274W} to properly account for the Poisson nature of the data without performing any background modeling.

\subsection{XMM-Newton and NuSTAR cross calibration}\label{sec:calib}

The \nustar observation only started 3 hours before the \xmm one came to an end. Considering the effect of solar flares in the \xmm observation (Figure~\ref{fig:cr_lc_xmm}), there is only an overlap of about 1 hour between MOS1/MOS2 and \nustar data to perform a joint X-ray analysis. If the \xmm spectrum has a high signal to noise ratio ($\sim1000\,\rm cts$ from the source in both MOS1 and MOS2 in 1 hour of observation), the count rates detected by \nustar imply that 1 hour of observation can only get to $\sim50$ source counts per FPM, resulting in very poor spectral constraints.

Nevertheless, we can use the overlapping times to ensure that the cross-calibration values between the detector modules in the two satellites is within the expected values from \citet{Madsen_2017}. 
We employ an absorbed, redshifted power-law model (\texttt{TBabs*zpowerlaw}), with absorption frozen to the Galactic value ($N_{\rm H}=1.34\times10^{21}\,\rm cm^{-2}$, \citealp{Willingale_2013}) using \citet{Wilms_2000} abundances. The instruments are linked via a cross normalization constant (\texttt{const}) that we keep frozen to 1 for FPMA. As for the priors, we use a Gaussian prior of mean 1 and standard deviation 0.05 for the FPMB constant, while we assign a Gaussian prior to the MOS1 constant with mean 1 and standard deviation 0.2, and we tie the MOS2 to the MOS1 constant. We use a uniform prior for the photon indices of FPMA and MOS1, and a log-uniform prior for the normalizations of FPMA and MOS1. The FPMB photon index is linked to that of FPMA, and the MOS2 photon index to that of MOS1. For the normalizations, we link the normalization of FPMB, MOS1 and MOS2 to those of FPMA. We also restrict our energy range to the overlapping one for the two instruments, $3-10\,\rm keV$. 

The results of the fit are as follows. The measured photon index is $\Gamma_{3-10\,\rm keV}=1.93_{-0.24}^{+0.30}$ for \nustar and $\Gamma_{3-10\,\rm keV}=2.16_{-0.33}^{+0.22}$ for \xmm, fully compatible with each other within $1\sigma$ uncertainties. As for the cross calibration constant, the FPMB value is $C_{\rm FPMB}=1.00_{-0.02}^{+0.07}$ and the MOS1 value is $C_{\rm MOS1}=0.87_{-0.16}^{+0.11}$, perfectly compatible with the expected 20\% difference between \xmm and \nustar instruments. Therefore, there are no additional instrumental biases that need to be accounted for in the rest of the analysis.

\begin{figure*}
    \centering
    \includegraphics[width=0.33\textwidth]{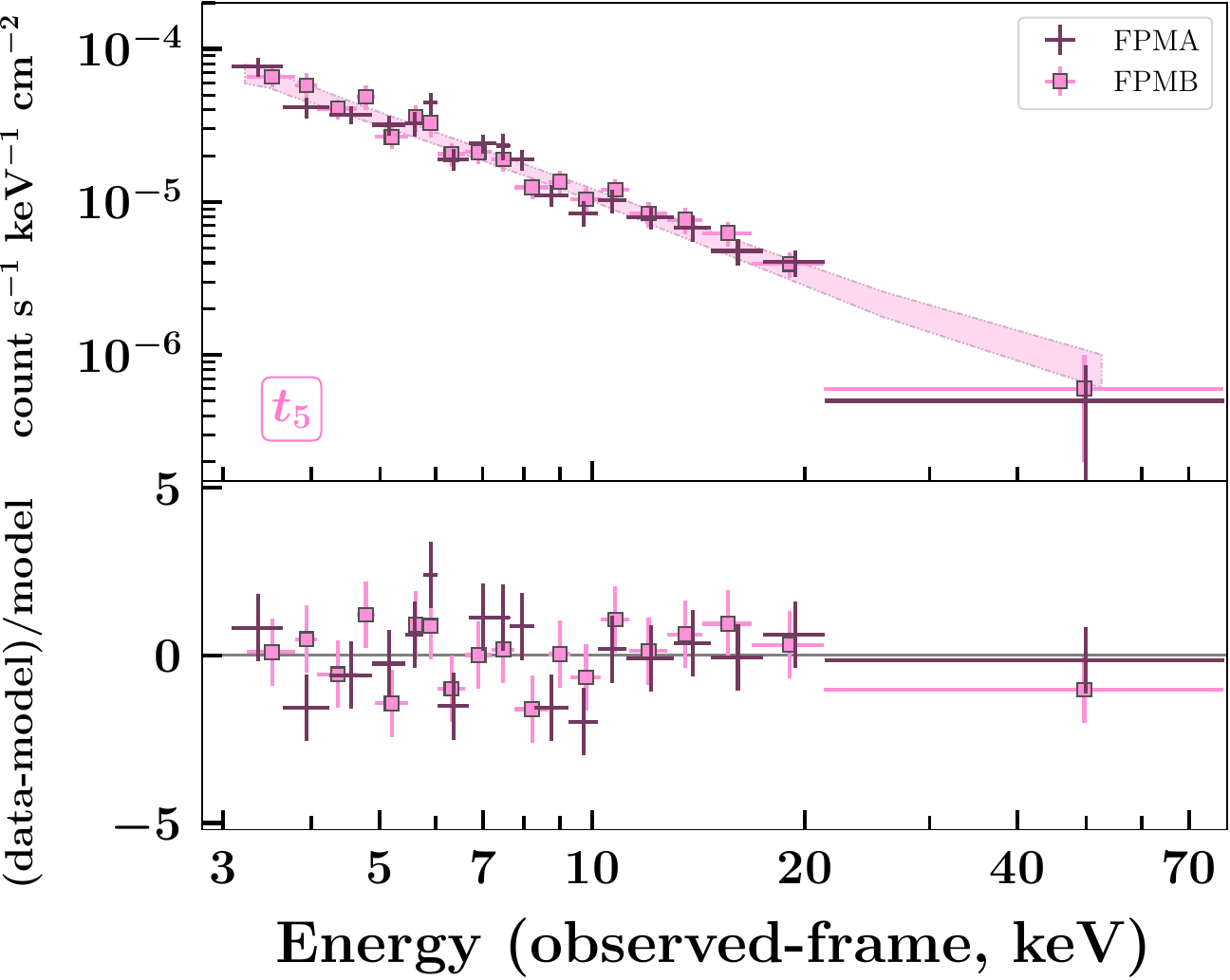}
    \includegraphics[width=0.33\textwidth]{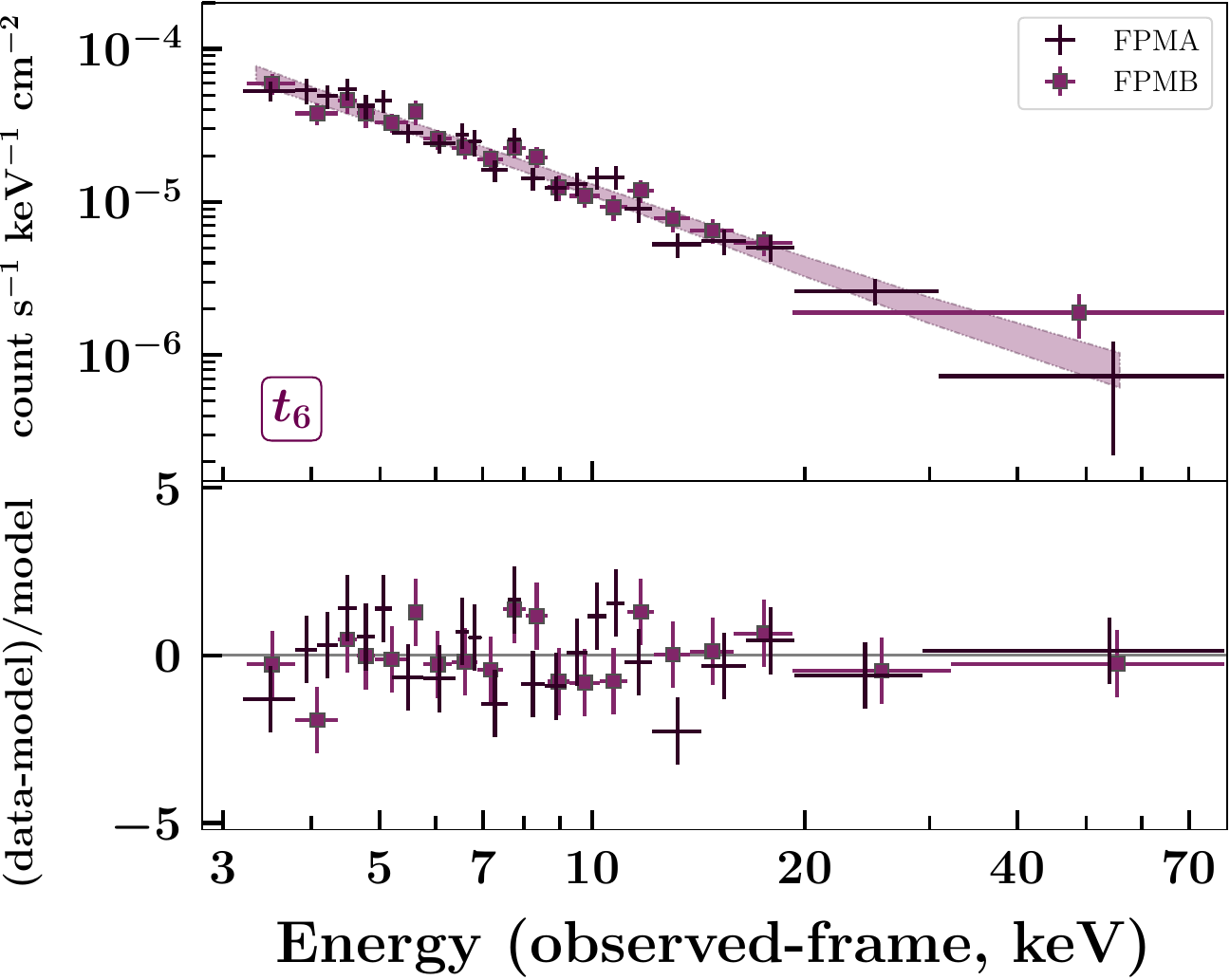}
    \includegraphics[width=0.33\textwidth]{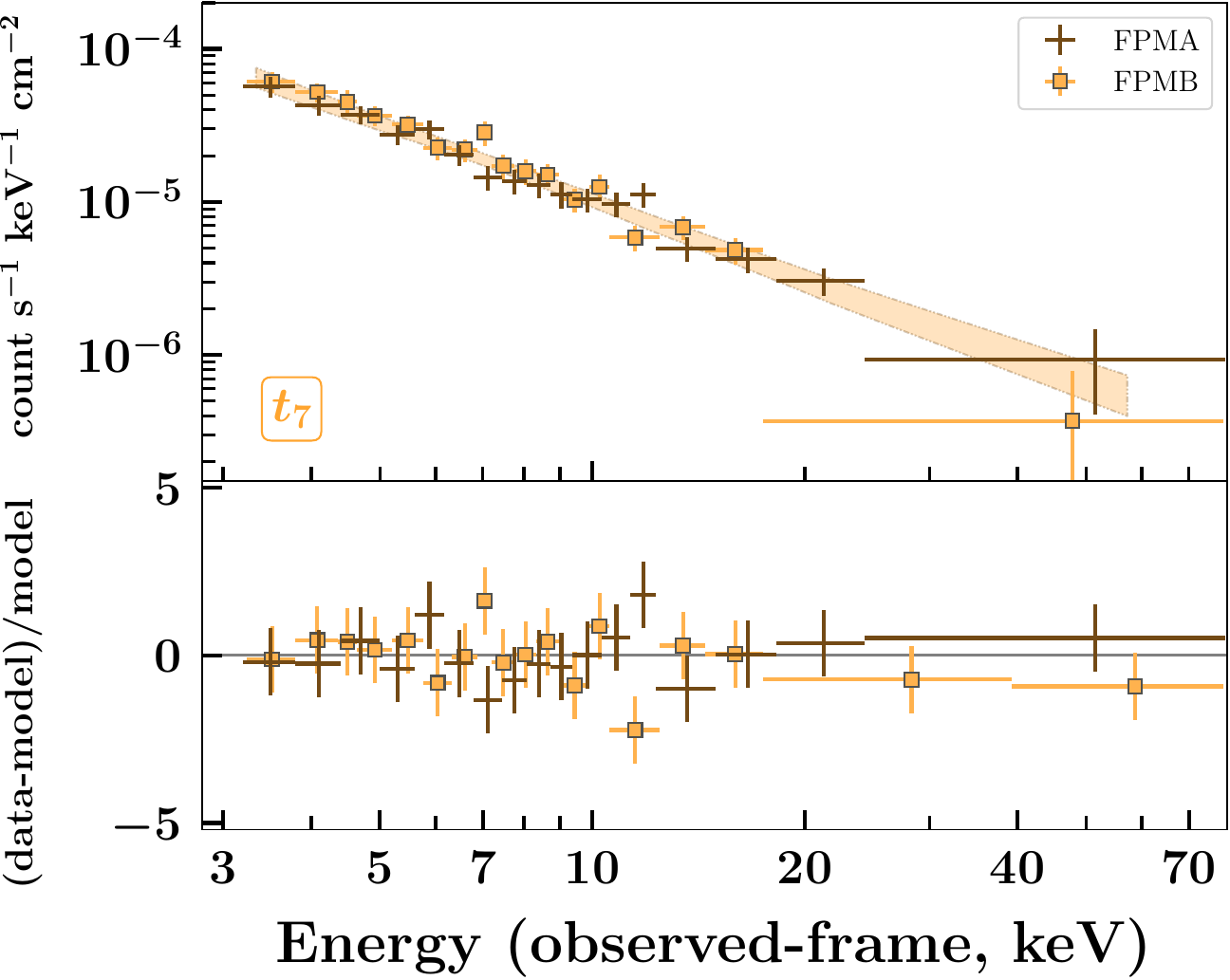}
    \includegraphics[width=0.33\textwidth]{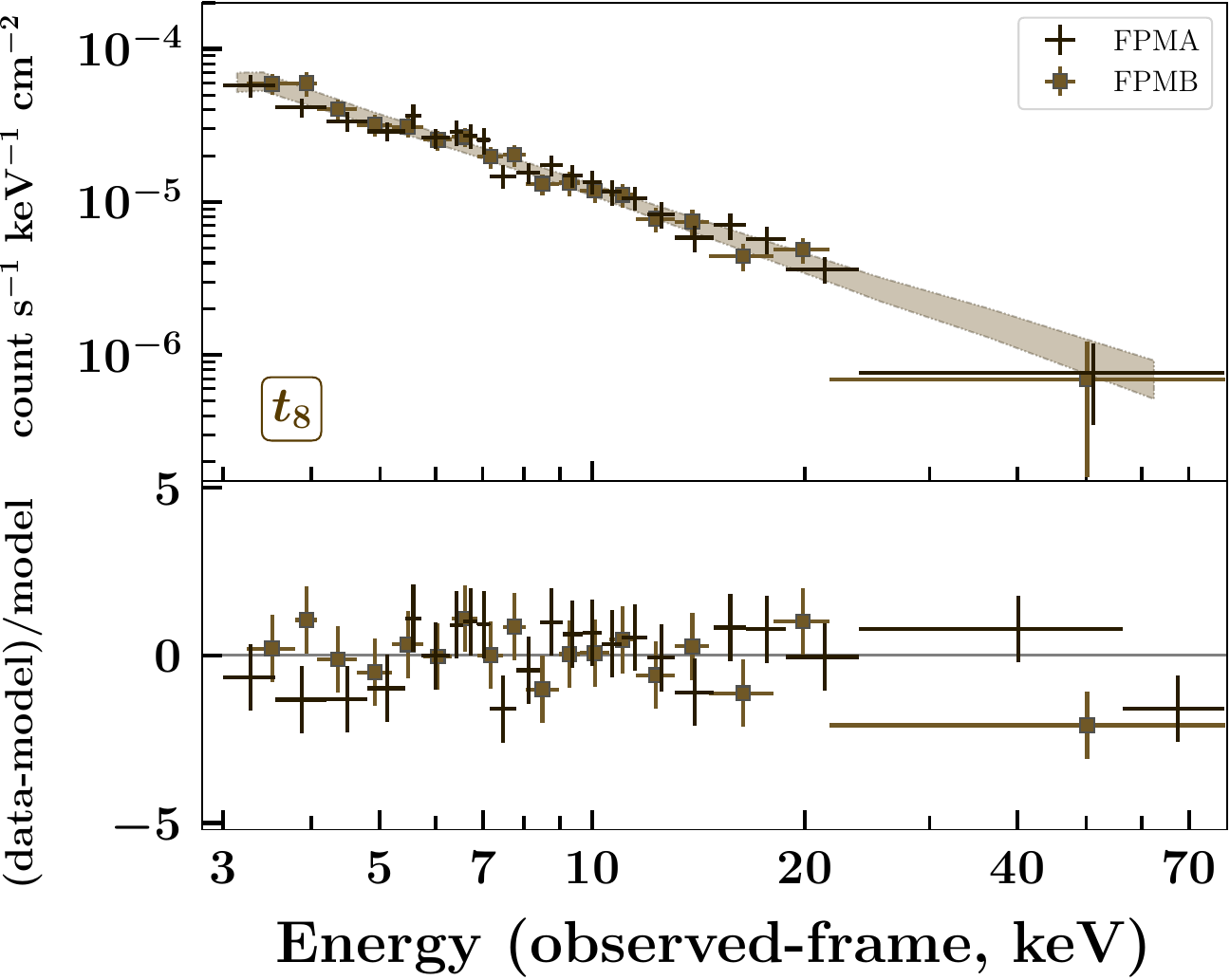}
    \includegraphics[width=0.33\textwidth]{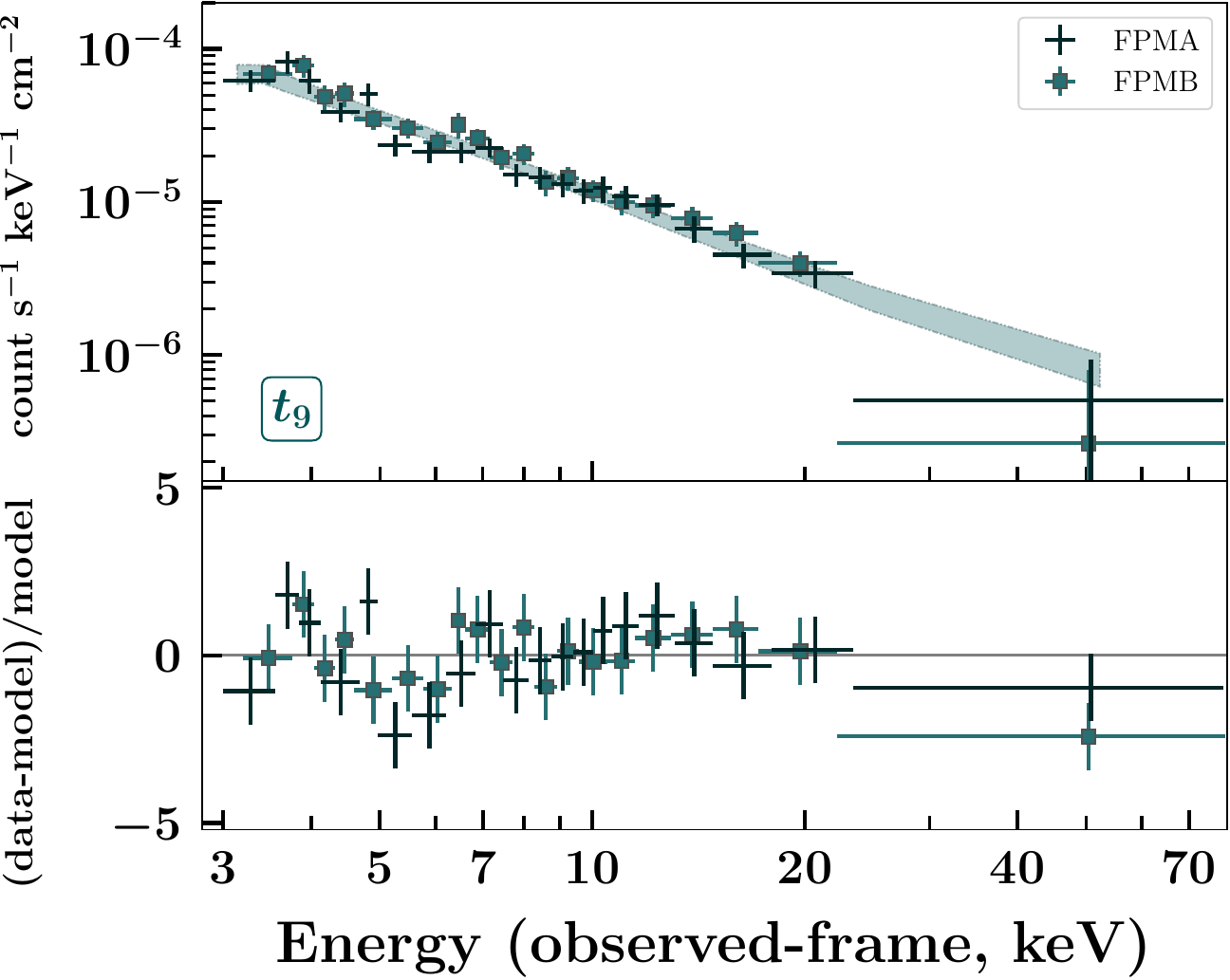}
    \includegraphics[width=0.33\textwidth]{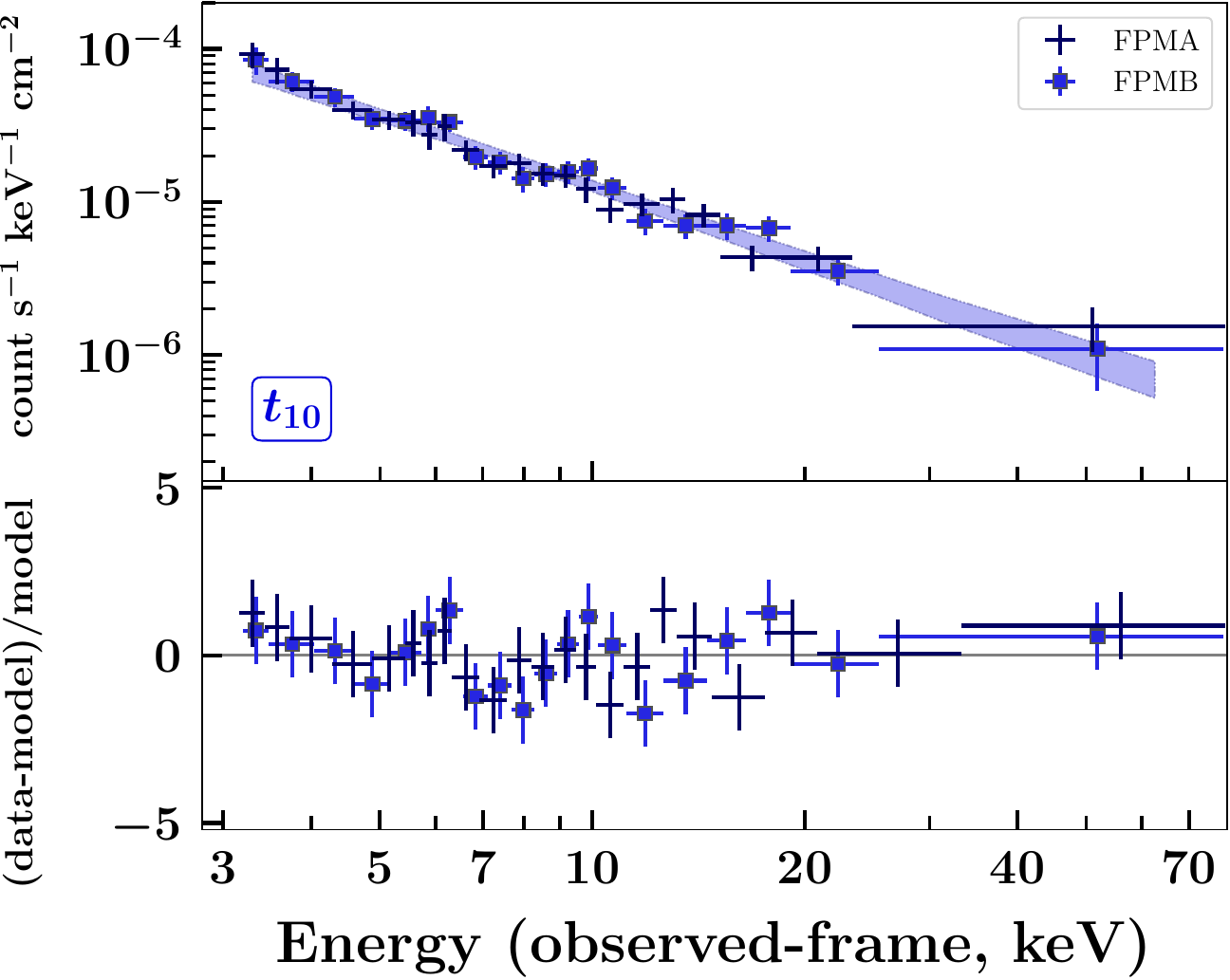}
    \includegraphics[width=0.33\textwidth]{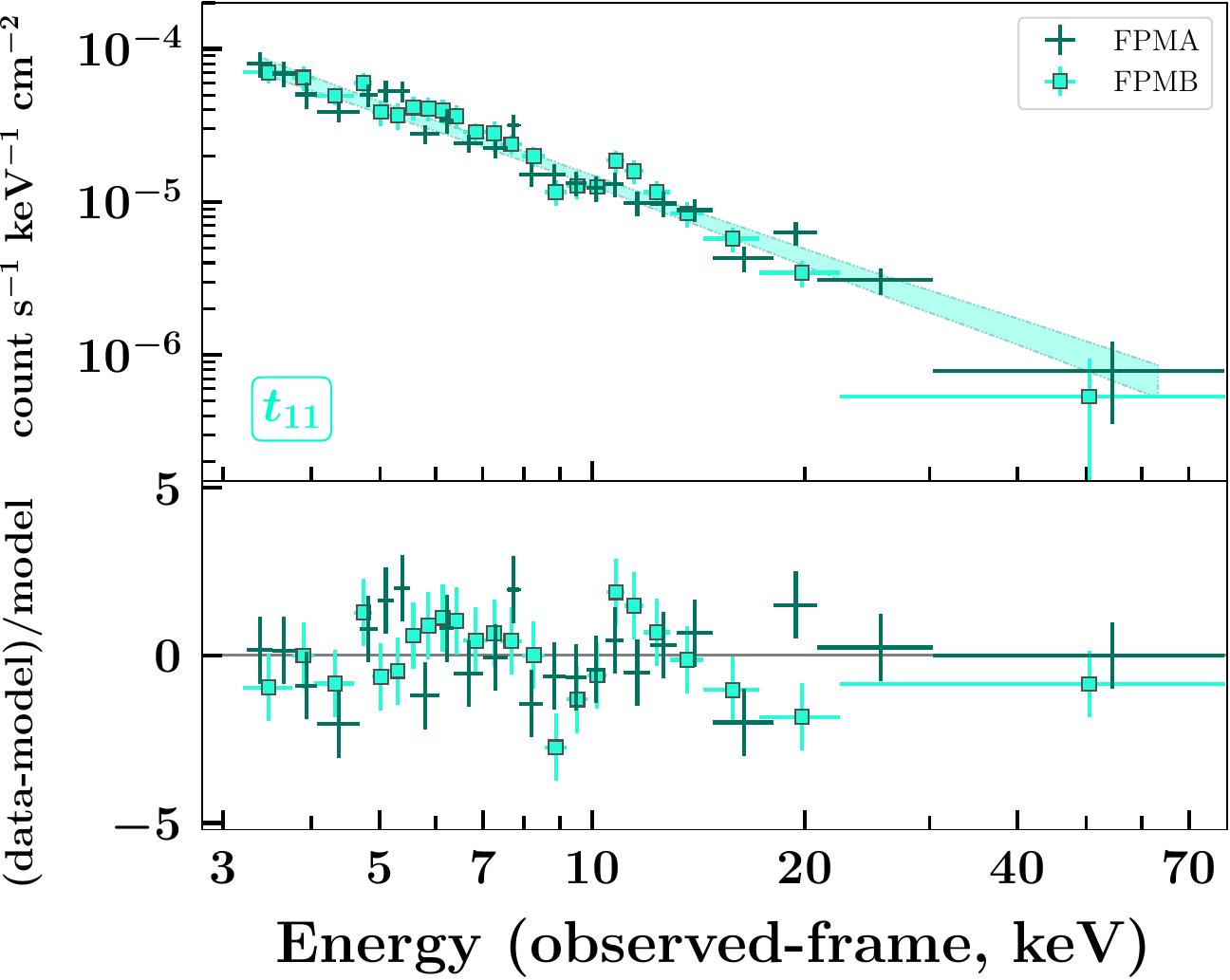}
    \includegraphics[width=0.33\textwidth]{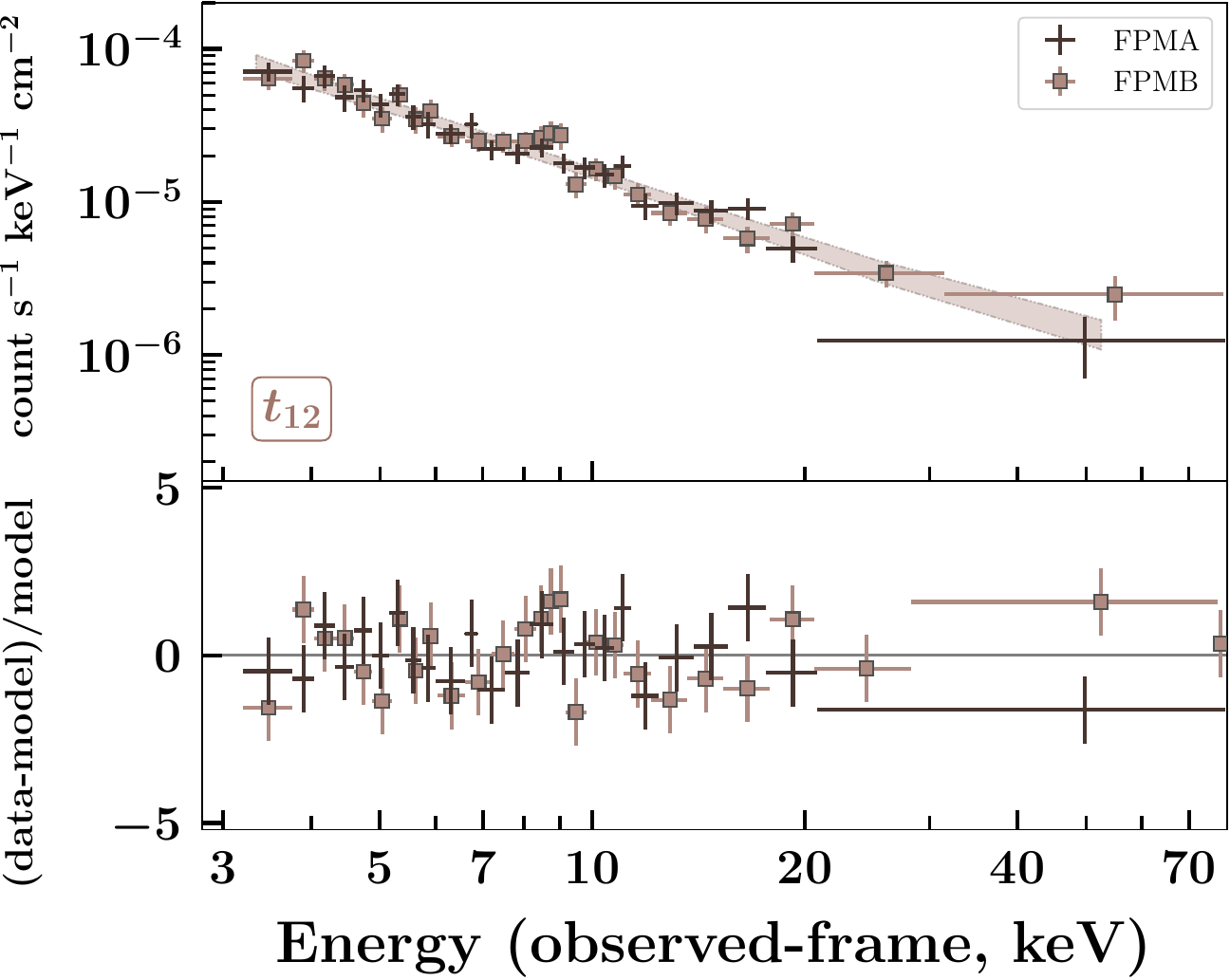}
    \includegraphics[width=0.33\textwidth]{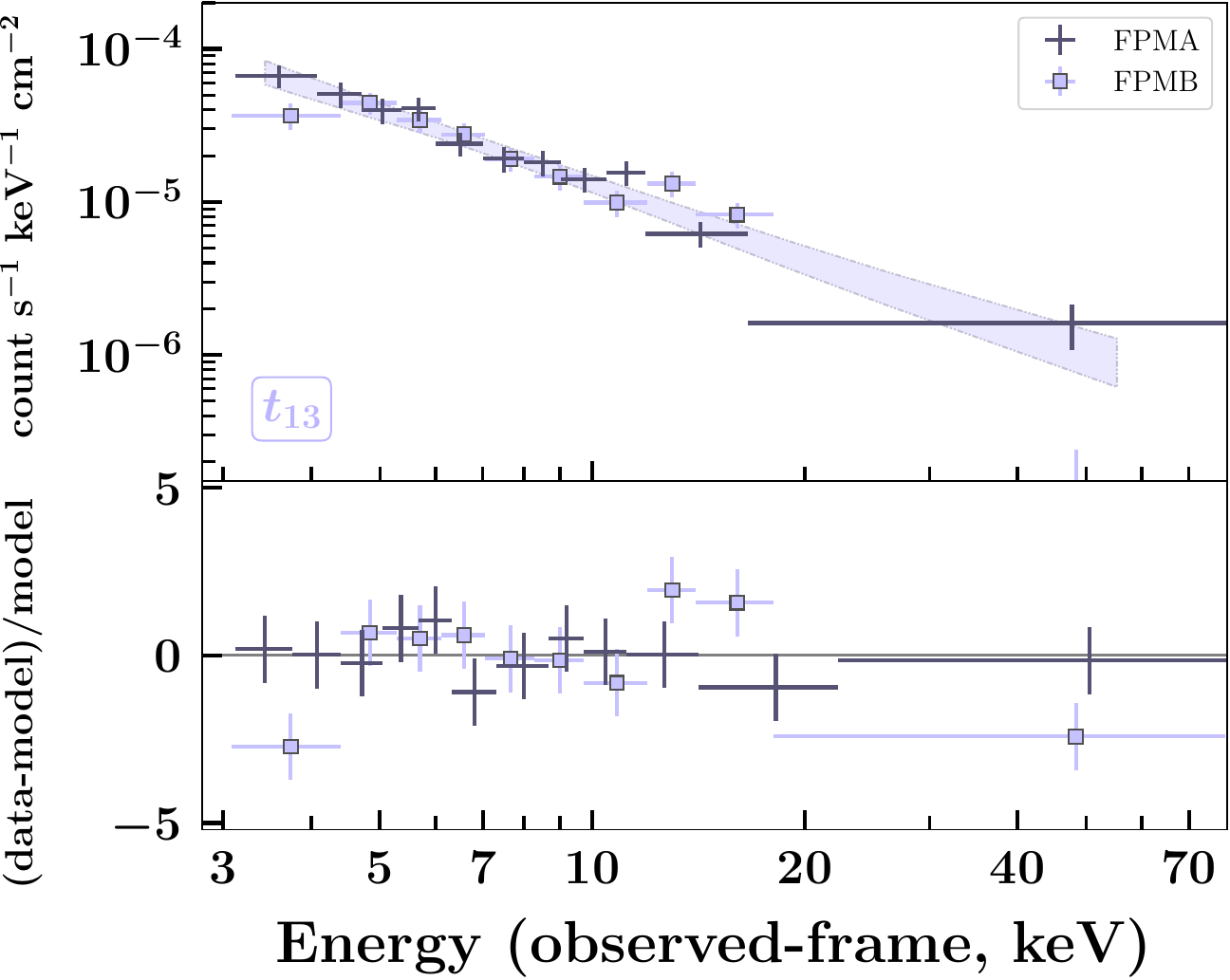}
    \caption{Background-subtracted \nustar count rate spectra and residuals, after binning each spectrum to have a minimum significance of five per bin for visual clarity. The source counts are normalized by the average effective area in each bin. The plots, from left to right, and top to bottom, are ordered as function of time bin, from $t_5$ to $t_{13}$ (see Table~\ref{tab:time_bins}) and the data points color coded from pink ($t_5$) to light purple ($t_{13}$). The FPMA and FPMB data are marked with crosses and squares, respectively. In each panel, we show the model predictions for the redshifted power-law fits to the data. Shaded areas (colored like the data points) are the 99th percentile range region of the total flux posterior distribution. }
    \label{fig:nu_pl_spec}
\end{figure*}
\begin{figure*}
    \centering
    \includegraphics[width=0.95\linewidth]{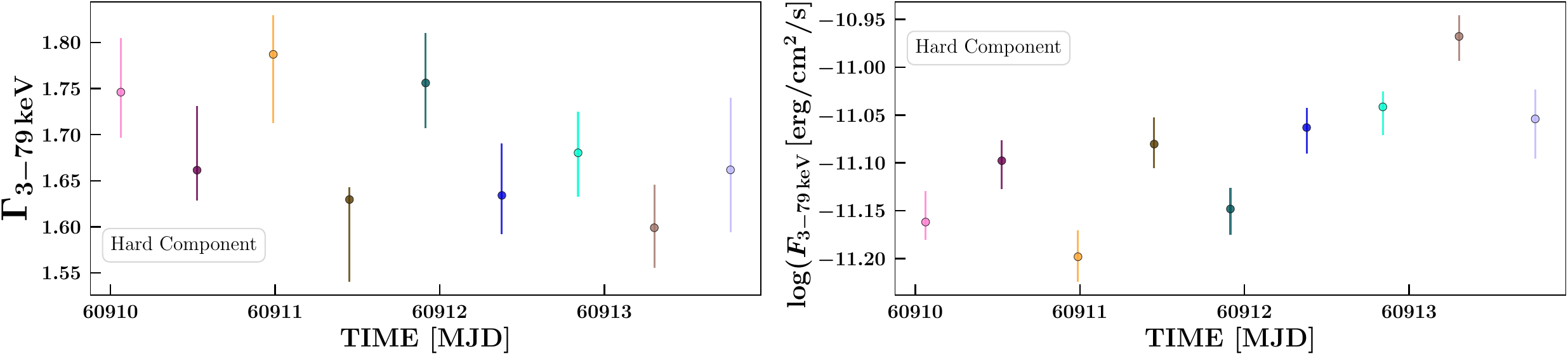}
    \caption{Temporal evolution of the $3-79\,\rm keV$ photon index (left) and flux (right) of the hard power-law component observed by \nustar.}
    \label{fig:ph_fl_nu}
\end{figure*}

\subsection{The XMM-Newton time lapse}

We started by fitting the broadband ($0.2-12\,\rm keV$) spectra of all \xmm epochs with a single absorbed redshifted power-law (\texttt{TBabs*zpowerlaw}), with absorption frozen to the Galactic value. For the first 4 epochs ($t_0$-$t_4$, see Table~\ref{tab:time_bins}) we fit MOS1, MOS2 and PN together, while, for the last epoch ($t_5$), we restrict our fit to only MOS1 and MOS2 due to the unavailability of PN data. The instruments are linked via a cross calibration constant frozen to 1 for MOS1 and free to vary for the others.  For the priors on the parameters, we use a Gaussian prior for the cross-calibration constants of MOS2 and PN with mean 1 and standard deviation of 0.05, a uniform prior for the photon index and a log-uniform prior for the normalization. The best-fit model parameters are reported in Table~\ref{tab:xmm_fit} and the spectral fits with the corresponding residuals are show in Figure~\ref{fig:xmm_fit_pl}. The results consistently return a best-fit photon index with $\Gamma\sim2$, which is quite soft with respect to typical FSRQ-like values in this energy range ($\Gamma_{\rm X}\sim1.5-1.8$). Importantly, looking at the spectral residuals of all epochs, we see a positive excess below 2 keV, suggesting that we require a second component to explain the \xmm data.

Therefore, we fit the same spectra at all epochs with two absorbed redshifted power-laws (\texttt{const*TBabs*(zpo+zpo)}). In this case, we use a uniform and log-uniform prior for both photon indices and normalizations, respectively. The instrumental constants are treated identically to the simple power-law case. The results of these fits are shown in Figure~\ref{fig:xmm_double_pl} and the values of the best fit parameters are reported in Table~\ref{tab:xmm_fit}.  
The light-curves of the derived photon indices and fluxes for the two components are shown in Figure~\ref{fig:xmm_lc_ph_fl}. As can be seen, this fit reveals two distinct spectral components: a hard power law with photon index $\Gamma\sim1.7 - 1.9$ (which we call the ``hard component''), and a soft component with $\Gamma\gtrsim2.5$ that emerges below approximately 2 keV (which we call the ``soft component''). Moreover, even if statistically there is very little evidence ($\gtrsim 1\sigma$) for a photon index or flux evolution of the two components, the modes of the distributions at different epochs suggest that the soft component becomes fainter and softer from $t_0$ to $t_5$.

To understand if the addition of this second component is preferred over the single power-law fit, we compute the Bayes factor and, following the Jeffreys scale \citep{Jeffreys_1939, Robert_2008}, we regard $\Delta\ln Z = \ln Z_1 - \ln Z_2>5$ as decisive evidence in favor of the model with the larger Bayesian evidence. For the first four epochs, the double power-law model is confidently preferred over the single power-law case ($\Delta \ln Z = 16-55$, see Table~\ref{tab:xmm_fit}). The last epoch instead shows a $\Delta \ln Z \sim 3.5$, which can only be considered suggestive evidence of the double power-law being favored over the single power-law case. This aligns with the fact that the ``soft component'' is getting fainter/softer by the end of the \xmm observation.

\subsection{The NuSTAR time lapse}

Since \nustar only samples energies above 3 keV and the ``soft component'' emerges below 2 keV, we fit the \nustar spectrum only with a single absorbed redshifted power-law (\texttt{TBabs*zpo}), with absorption frozen at the Galactic value. The priors choice is a uniform/log-uniform distribution for the photon index/normalization, and a gaussian prior for the FPMB constant centered at 1 with 0.05 standard deviation. The fit is performed over the full $3-79\,\rm keV$ band. The results of these fits are shown in Figure~\ref{fig:nu_pl_spec} and the values of the best fit parameters are reported in Table~\ref{tab:nu_fit}.  The light-curves of the derived photon indices and flux for the component are shown in Figure~\ref{fig:ph_fl_nu}. The photon index, which is tracing the evolution of the ``hard component'', is between $\sim 1.6-1.8$, remaining compatible at all epochs within $1\sigma$ uncertainties. 
Instead, what is statistically significant is an increase in flux towards the end of the observation ($t_{12}$) of a factor of 1.5 with respect to $t_5$. In the last time bin ($t_{13}$) the flux appears to start decreasing again.

\section{Results \& Interpretation}\label{sec:res_disc}

\begin{figure*}[t]
    \centering
    \includegraphics[width=0.97\linewidth]{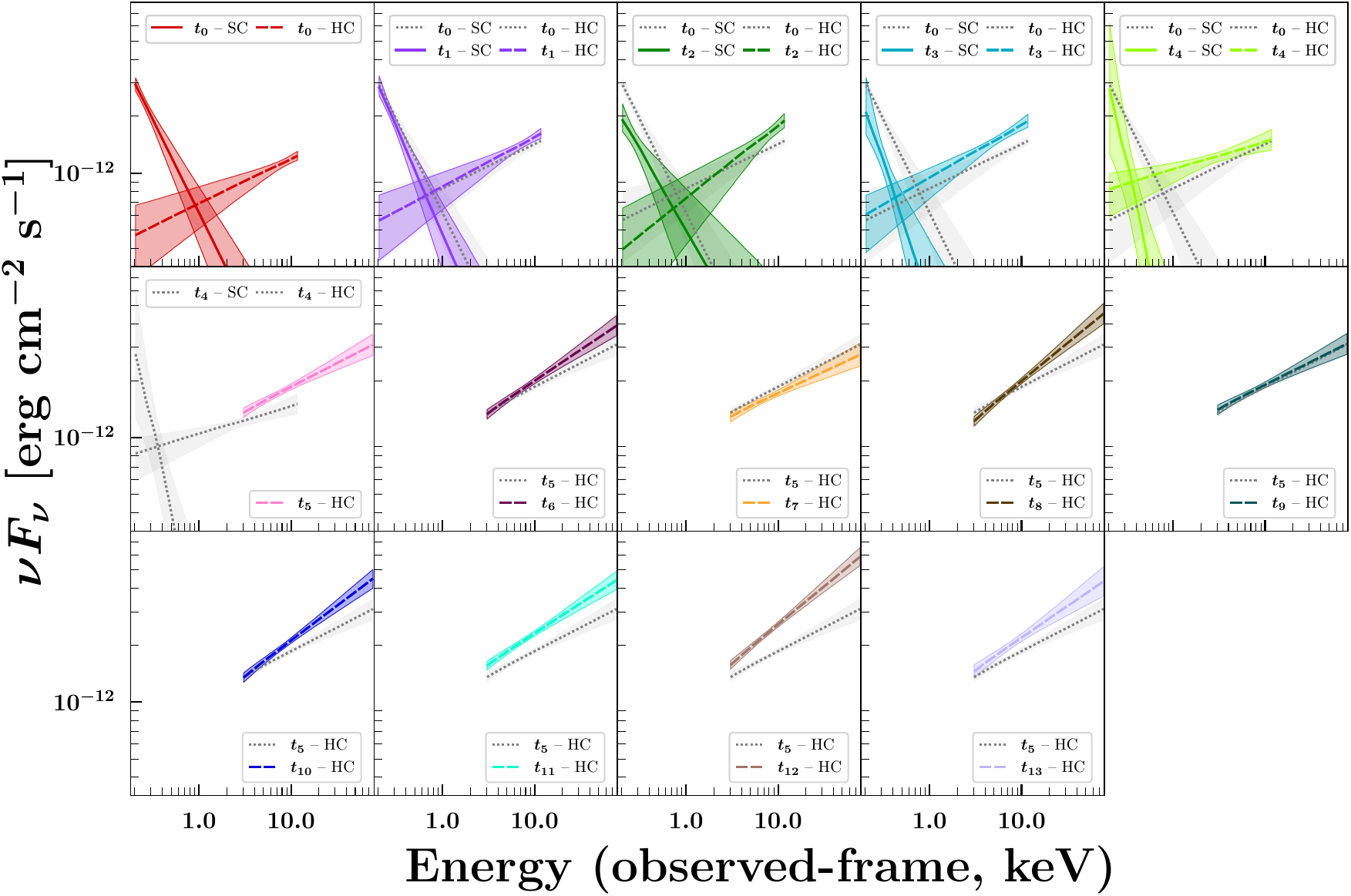}
    \caption{ Broadband X-ray SED (0.2-79 keV) of the target and its evolution as function of time, from the start of the \xmm observation ($t_{0}$, red, upper left panel) to the end of the \nustar one ($t_{13}$, light purple, bottom mid panel). The solid line represent the soft component (SC), while dashed line the hard component (HC) of the spectrum. The lines are the mode of the flux posteriors while the bands represent the 68\% HDIs. The \xmm SEDs have been multiplied by a factor of 1.2 to account for the 20\% instrumental difference in recorded flux between the two instruments (see Section~\ref{sec:calib}). At different times, we also show with the gray shaded region and dotted lines a reference SED: from $t_1$ to $t_4$, we use the SED at $t_0$ of the soft and hard component; at $t_5$ we compare the SED of the hard component sampled by \nustar with the $t_4$ soft+hard component sampled by \xmm; finally from $t_6$ to $t_{13}$ we show the $t_5$ SED of the hard component. As time goes on, the soft component moves out of the soft X-ray band, while the hard component brightens, following the shift of the two SED components to lower energies as the flare progresses.}
    \label{fig:sed_time}
\end{figure*}

The X-ray spectral timing analysis of this unprecedented monitoring of the days after the VHE flare of PKS 1725+123 enables us to follow the SED evolution of the source with  exquisite detail. To visually combine and interpret the results of our campaign, in Figure~\ref{fig:sed_time} we show the time evolution of the $0.2-79\,\rm keV$ SED component in the $\nu F_\nu$ space. The shaded regions trace the 68\% HDI of the flux posteriors of the fit at different times, from $t_0$ to $t_{13}$ (see Table~\ref{tab:time_bins}). 

At the start of the \xmm campaign (red region), both the soft and hard components of the SED are present. As time goes by, the soft component gets fainter and softer, shifting out of the X-ray band ($t_4$, light green). 
This effect is visible in a timescale of just $\sim 14$ hours (observed frame). Then, \nustar starts monitoring the source. Within 4 days, the hard component brightened by a factor of $\sim 1.5$ ($t_{12}$, brown) to then decrease in flux in the last time bin ($t_{13}$, light purple).      

This SED evolution can be phenomenologically interpreted as follows. First of, the presence of the two spectral components detected with \xmm likely marks the transition between two emission processes in the jet of PKS 1725+123, such as the synchrotron component of the SED and the high-energy one, that is, the ``valley" of the SED \citep[see also][]{wierzcholskawagner16}, or the synchrotron emission from two emission zones \citep[e.g.,][]{Liu_2026}. It is unlikely that this transition could be explained by a curved continuum produced by a single electron population in the jet, as the V-shaped spectrum would require the electron distribution to harden only at its highest energies.
As time goes on, the tail-end of the synchrotron component shifts out of the X-ray band, while the high-energy component of the SED shifts back in. Taken together, these results could suggest that we are witnessing the shift of the SED peaks to lower energies in a matter of days.

Putting these results into context with the VHE flare that happened a few days prior to  the X-ray monitoring, phenomenologically one can interpret that a physical process (shock, magnetic reconnection, or other) happened in or around the jet to enable particles to be accelerated up to the VHE regime. Assuming the broadband SED throughout the flare could be explained by a one-zone jet model \citep[see][for a 2-zone interpretation]{Liu_2026}, both the synchrotron and the high-energy SED peaks brightened and shifted to the highest energies, and by doing so PKS~1725+123 temporarily transitioned into an ISP blazar. As the flare progressed and energy dissipated, the SED peaks started to return to their ``normal'' state (i.e., the typical LSP state of PKS 1725+123), shifting to lower energies and later decreasing in brightness, hence explaining the evolution of the X-ray spectrum. 
The Fermi-LAT results presented in \citet{Liu_2026} also seem to agree with this interpretation, as they show that during the VHE flare the LAT $\gamma$-ray photon index hardened ($\Gamma_{\gamma}\sim1.7$) and the source brightened, about a week later around the time of our X-ray campaign the $\gamma$-ray flux decreased by a factor of $\sim4$ and the photon index softened ($\Gamma_{\gamma}>2$).  
For the first time, we are witnessing in X-rays the retraction of the SED towards its normal state in real time, and we find that in the observed time frame this took about $\sim1$ week from the VHE discovery. Full SED modeling of the source is reserved for a future publication, and inference on the intrinsic variability timescales will be strongly dependent on the derived jet parameters and inferred jet geometry. However, employing the values of the bulk Lorentz factors reported in the recent analysis of the flare by \citet[][$\Gamma=35$ for the compact zone, and $\Gamma=5.9$ for the extended zone, see also \citealp{Chaudhary_2026}]{Liu_2026}, assuming a jet viewing angle of the jet $\theta_{\rm v}\sim1/\Gamma$ and, hence, $\delta\sim\Gamma$, this implies that the intrinsic timescales of variability are $t_{\rm intr}/t_{\rm obs} = \delta/(1+z)\sim 4-22$. If these parameters are representative of the jet of PKS 1725+123, the intrinsic time scale in which the jet went from an ISP back to an LSP like object is in the order of $\sim 4-22$ weeks, therefore the region(s) responsible for this enhancement are $R\leq ct_{\rm intr} \sim 7\times10^{16}-4\times10^{17}\rm cm$ 
in size. 
Depending on the boosting parameters which a complete time-resolved broadband SED modeling will give us, we will be able to place stringent constraints on the size and location of the emission region, on the acceleration mechanisms at play, and on the shape and evolution of the particle energy distribution.

\section{Conclusions and Outlook}\label{sec:outl}

With nearly five days of continuous soft-to-hard X-ray coverage following the VHE flare of PKS 1725+123, this campaign offers a unique opportunity to trace the day-scale evolution of the X-ray emission of PKS 1725+123 in exceptional detail.
Our results show that \xmm caught the presence of the synchrotron component of the SED receding from a dramatic shift, which had moved its peak into the UV band at the beginning of the VHE flare, and then started receding back to the ``normal'' LSP state of the source with a peak in the mid-infrared range \citep[e.g., Figure 6 of][]{Liu_2026}. 
On the other hand, \nustar caught the evolution of the high energy SED component. 
Since the onset of the VHE flare, other works have followed the X-ray evolution of the source \citep{Liu_2026}. However, the presence of these two SED components could not be distinguished due to lower spectral sensitivity and energy coverage of their instruments.

This \xmm and \nustar X-ray campaign results will be crucial to understand the full SED evolution of the target and to probe the physical processes that governed the jet in such an enhanced state. At the time of the VHE flare, our team also collected a dense multi-wavelength campaign coordinating simultaneous observations spanning the radio to the VHE $\gamma$-ray bands. Together with available multi-wavelength datasets pre- and post-flare, these data will allow us to place the VHE flare in the context of the long-term behavior of PKS 1725+123 relativistic jet. 
SED modeling will be used to investigate the physical conditions driving the flare, while variability correlations across wavebands will help disentangle the different emission components. These efforts will ultimately shed light on the particle acceleration processes that power this blazar jet and if a one-zone or multiple-zone model is needed to explain the multi-wavelength data (publication in preparation).
As facilities such as the Cherenkov Telescope Array Observatory (CTAO) will come online, the possibility of catching VHE flares from even higher redshift blazars will increase. Therefore, coordinated simultaneous campaigns with X-ray facilities would be extremely beneficial to truly understand the acceleration and radiation mechanisms on these powerful jets. 

\section*{Acknowledgments}
The team would like to thank both the \xmm and the \nustar SOC for working with us in scheduling this campaign as timely as possible. LM acknowledges that this work was supported by the Initiative and Networking Fund of the Helmholtz Association under the Helmholtz Investigator Groups Programme, call 2025 (VH-NG-21-01). LM and AG acknowledge support from DESY (Zeuthen, Germany), a member of the Helmholtz Association HGF.
M.Z. acknowledges funding by the Deutsche Forschungsgemeinschaft (DFG, German Research Foundation) -- project number 460248186 (PUNCH4NFDI).
L.H. acknowledges that this work benefited from the support of the project COCOA-NuGETs ANR-23-CE31-0026 of the French National Research Agency (ANR).

\begin{table*}[t]
\centering
\small
   \caption{\xmm~$0.2-12\,\rm keV$ spectral results for all time bins. The preferred model based on the Bayes factor is marked in boldface.} 
   \begin{tabular}{ccccccccc}
    Time bin & Model & $\Gamma_1$ & $\log(N_1)$ & $\Gamma_2$ & $\log(N_2)$ & $C_{\rm MOS2}$  & $C_{\rm PN}$ & $\ln Z$  \\   
     &  &  & $\rm [ph~cm^{-2}~s^{-1}~keV^{-1}]$ &  & $\rm [ph~cm^{-2}~s^{-1}~keV^{-1}]$ &  &  &  \\       
\hline
$t_0$ & zpow & $2.18_{-0.02}^{+0.01}$ & $-2.67_{-0.01}^{+0.01}$ & -- & -- & $1.06_{-0.03}^{+0.01}$ & $1.09_{-0.01}^{+0.03}$ & $-231.21$ \\
   & {\bf zpow+zpow} & $1.79_{-0.13}^{+0.09}$ & $-2.97_{-0.13}^{+0.11}$ & $2.87_{-0.17}^{+0.30}$ & $-2.86_{-0.13}^{+0.08}$ & $1.06_{-0.02}^{+0.03}$ & $1.07_{-0.02}^{+0.02}$ & $-175.37$ \\   
\hline
$t_1$ & zpow & $2.11_{-0.02}^{+0.01}$ & $-2.72_{-0.01}^{+0.01}$ & -- & -- & $0.96_{-0.02}^{+0.02}$ & $1.04_{-0.01}^{+0.03}$ & $-233.06$ \\
  & {\bf zpow+zpow} & $1.78_{-0.14}^{+0.07}$ &  $-2.98_{-0.13}^{+0.09}$ & $3.05_{-0.29}^{+0.32}$ &  $-2.99_{-0.11}^{+0.11}$  & $0.97_{-0.03}^{+0.02}$ & $1.03_{-0.01}^{+0.03}$ & $-175.36$  \\ 
\hline
$t_2$ & zpow & $1.97_{-0.02}^{+0.02}$ & $-2.78_{-0.01}^{+0.01}$ & -- & --  & $0.98_{-0.02}^{+0.03}$ & $1.06_{-0.04}^{+0.02}$ & $-191.56$ \\
  & {\bf zpow+zpow} & $1.67_{-0.20}^{+0.13}$ & $-2.97_{-0.28}^{+0.08}$ & $2.54_{-0.21}^{+0.60}$ &  $-2.94_{-0.18}^{+0.15}$  & $0.99_{-0.02}^{+0.02}$ & $1.04_{-0.03}^{+0.03}$ & $-175.49$  \\ 
\hline
$t_3$ & zpow & $1.93_{-0.02}^{+0.01}$ & $-2.82_{-0.01}^{+0.01}$ & -- & -- & $1.02_{-0.03}^{+0.01}$ & $1.08_{-0.04}^{+0.02}$ & $-165.62$ \\
  & {\bf zpow+zpow} & $1.76_{-0.11}^{+0.07}$ & $-2.91_{-0.10}^{+0.05}$ & $3.60_{-0.91}^{+0.35}$ & $-3.28_{-0.19}^{+0.22}$  & $1.01_{-0.03}^{+0.02}$ & $1.05_{-0.03}^{+0.03}$ & $-148.47$  \\ 
\hline
$t_4$ & zpow & $1.98_{-0.03}^{+0.04}$ & $-2.82_{-0.01}^{+0.02}$ & -- & -- & $1.00_{-0.01}^{+0.05}$ & -- & $-130.44$ \\
  & {\bf zpow+zpow} & $1.87_{-0.10}^{+0.08}$ & $-2.87_{-0.06}^{+0.05}$ & $4.29_{-0.97}^{+1.17}$ & $-3.58_{-0.24}^{+0.38}$ & $1.01_{-0.02}^{+0.04}$ & -- & $-127.09$  \\
\hline
   \end{tabular}\label{tab:xmm_fit}
\end{table*}

\begin{table*}[t]
\centering
   \caption{\nustar $3-79\,\rm keV$ spectral results for all time bins.} 
\begin{tabular}{ccccc}
    Time bin & Model & $\Gamma$ & $\log(N)$ 
    & $C_{\rm FPMB}$ \\ 
     &   &   & $\rm [ph~cm^{-2}~s^{-1}~keV^{-1}]$ 
    &   \\   
\hline
$t_5$ & zpow & $1.75_{-0.05}^{+0.06}$ &  $-2.86_{-0.04}^{+0.08}$ & $1.00_{-0.04}^{+0.04}$  \\
$t_6$ & zpow & $1.66_{-0.03}^{+0.07}$  & $-2.89_{-0.06}^{+0.06}$ & $0.99_{-0.02}^{+0.05}$ \\
$t_7$ & zpow &  $1.79_{-0.07}^{+0.04}$ & $-2.84_{-0.07}^{+0.06}$ & $1.03_{-0.03}^{+0.04}$  \\
$t_8$ & zpow & $1.63_{-0.09}^{+0.01}$  & $-2.97_{-0.07}^{+0.04}$ & $0.99_{-0.03}^{+0.04}$ \\
$t_9$ & zpow & $1.76_{-0.05}^{+0.05}$ & $-2.81_{-0.06}^{+0.06}$ & $1.04_{-0.03}^{+0.05}$  \\
$t_{10}$ & zpow & $1.63_{-0.04}^{+0.06}$ & $-2.90_{-0.07}^{+0.04}$ & $1.01_{-0.02}^{+0.05}$  \\
$t_{11}$ & zpow & $1.68_{-0.05}^{+0.04}$ & $-2.82_{-0.06}^{+0.04}$ & $1.03_{-0.03}^{+0.04}$  \\
$t_{12}$ & zpow & $1.60_{-0.04}^{+0.05}$  & $-2.89_{-0.04}^{+0.06}$ & $1.00_{-0.02}^{+0.05}$ \\
$t_{13}$ & zpow & $1.66_{-0.07}^{+0.08}$  & $-2.87_{-0.08}^{+0.08}$ & $0.97_{-0.03}^{+0.05}$ \\
\hline
\end{tabular}\label{tab:nu_fit}
\end{table*}

\appendix

\section{XMM-Newton and NuSTAR GTIs}\label{app:a}

In Table~\ref{tab:time_bins} we list the time bins employed for the X-ray spectral timing analysis of \xmm and \nustar data.

\begin{table}[h]
    \centering
\begin{threeparttable}    
    \caption{Time bins employed for the spectral timing analysis of \xmm ($\Delta t =10\,\rm ks$) and \nustar ($\Delta t =40\,\rm ks$). }
    \begin{tabular}{cccc}
     Instrument &  Id. & TSTAR$^{a}$ & TSTOP$^{a}$ \\   
                & & $[10^8\,\rm s]$ & $[10^8\,\rm s]$  \\   
    \hline
    \xmm      &  $t_0$ & 8.72229469 & 8.72241774 \\
              & $t_1$ & 8.72241774 & 8.72254079  \\
              & $t_2$ & 8.72254079 & 8.72266383 \\
              & $t_3$ & 8.72266383 & 8.72278688 \\
              & $t_4$ & 8.72278688 & 8.72290993 \\
    \hline
    \nustar    & $t_5$ & 4.9358855 & 4.9362855  \\
               & $t_6$ & 4.9362855 & 4.9366855  \\
               & $t_7$ & 4.9366855 & 4.9370855 \\
               & $t_8$ & 4.9370855 & 4.9374855 \\
               & $t_9$ & 4.9374855 & 4.9378855 \\
               & $t_{10}$ & 4.9378855 & 4.9382855 \\
               & $t_{11}$ & 4.9382855 & 4.9386855 \\
               & $t_{12}$ & 4.9386855 & 4.9390855 \\
               & $t_{13}$ & 4.9390855 & 4.9392359\\
    \hline
    \end{tabular}
    \begin{tablenotes}
      \footnotesize
      \item[$a$] For \xmm seconds, the reference time is January 1, 1998, at 00:00:00 TT; for \nustar seconds, the referce time is January 1, 2010 at 00:00:00 UTC.
    \end{tablenotes}
  \end{threeparttable}    
    \label{tab:time_bins}
\end{table}

\section{Boundaries of fit parameters}\label{app:b}

To allow fit reproducibility, we list in Table~\ref{tab:params_prior} the parameter ranges and associated priors for all X-ray models employed in the paper.

\begin{table}[h]
    \centering
    \small
    \caption{List of parameter ranges and priors employed in the fits.}
    \begin{tabular}{cccc}
    Model & Params Name  & Params Range & Type of Prior    \\
       \hline
       zpow   &    $\Gamma$    & [$0,6$] & uniform \\
         &   $ N $   & [$10^{-6}$, $10^{-1}$] & log-uniform\\
       \hline
        zpow+zpow   &   $\Gamma_1$    & [$0,2.5$] & uniform \\
          &  $N_1$   & [$10^{-6}$, $10^{-1}$] & log-uniform\\
          &  $\Gamma_2$    & [2,6] & uniform \\
          &  $N_2$   & [$10^{-6}$, $10^{-1}$] & log-uniform\\
       \hline
       zbknpow &   $\Gamma_1$ & [$0,2.5$] & uniform \\
          & $E_b$      & [$10^{-3}$, $10^{2}$] & log-uniform \\
          & $\Gamma_2$    & [$2,6$] & uniform \\
          & $N$   & [$10^{-6}$, $10^{-1}$] & log-uniform\\
       \hline
       const &   $C$ & [$0.1,10$] & Gaussian/uniform \\       
    \end{tabular}
    
    \label{tab:params_prior}
\end{table}

\bibliography{pks.bib}{}
\bibliographystyle{aasjournalv7}

\end{document}